\documentclass[review,fleqn,sort&compress]{elsarticle}
\usepackage{amssymb,amsbsy,amsthm,amsmath,amsfonts,amssymb,amscd}
\usepackage{geometry}
\usepackage{babel,blindtext}
\usepackage{lipsum}
\usepackage{bm} 
\usepackage{graphicx}
\usepackage{tabularx}
\graphicspath{{./images/}}
\usepackage{float}
\usepackage{caption}
\usepackage{subcaption}
\usepackage{multirow}
\usepackage{cancel}
\usepackage{lineno}
\usepackage{hyperref}
\usepackage{graphicx}
\usepackage{mathrsfs}
\usepackage{booktabs}
\hypersetup{colorlinks, citecolor=black, filecolor=black, linkcolor=black, urlcolor=black}
\usepackage{booktabs}
\usepackage[ruled,linesnumbered]{algorithm2e}
\usepackage[toc,page,titletoc]{appendix}
\usepackage[capitalise,nameinlink]{cleveref}
\crefname{supp}{Supplement}{Supplements}
\usepackage{algpseudocode}
\usepackage{mathtools,nccmath}
\usepackage{adjustbox}
\usepackage{cleveref}
\usepackage{xparse}
\theoremstyle{definition}

\usepackage{color}
\usepackage{xcolor}
\usepackage{amsmath}
\usepackage{soul}
\colorlet{soulred}{red!40}

\definecolor{airforceblue}{rgb}{0.36, 0.54, 0.66}

\begin{document}
\begin{frontmatter}
\title{Modified non-local damage model: resolving spurious damage evolution}




\author{Roshan Philip Saji ${}^{1,2}$}
\author{Panos Pantidis ${}^3$}
\author{Mostafa E. Mobasher ${}^{1,3}$\corref{cor1}}

\cortext[cor1]{Corresponding author. \emph{E-mail address:} \texttt{mostafa.mobasher@nyu.edu} (Mostafa E. Mobasher)}
\address{${}^1$ Mechanical Engineering Department, Tandon School of Engineering, New York University,  6 MetroTech Center, Brooklyn, NY 11201, USA}
\address{${}^2$ Mechanical Engineering Department, New York University Abu Dhabi, Abu Dhabi, P.O. Box 129188, UAE}
\address{${}^3$ Civil and Urban Engineering Department, New York University Abu Dhabi, Abu Dhabi, P.O. Box 129188, UAE}

\begin{abstract}
Accurate prediction of damage and fracture evolution is critical for the safety design and preventive maintenance of engineering structures, however existing computational methods face significant limitations. On one hand, discrete damage and phase-field models are often computationally prohibitive for real world applications and they are less generalizable across different material classes. On the other hand, conventional gradient damage models which are based on phenomenological laws, though more computationally efficient, they suffer from unrealistic widening of the damage-band as damage progresses. 

This paper presents a modified non-local gradient damage model (MNLD) that overcomes these shortcomings by introducing modifications to the stress degradation function and forcing term in the Helmholtz free energy expression. These two modifications ensure that as damage approaches its maximum value, both the thermodynamic damage driving force for damage vanishes and the evolution of the forcing term decays. Consequently, the damage band retains a non-growing constant width throughout its evolution. The proposed approach builds on insights gained from two intermediate models, which addressed the necessary conditions separately before integrating them into a unified formulation. Numerical validation is performed on several 1D and 2D benchmark problems, demonstrating that the proposed model can reliably produce fixed-width damage bands. The proposed approach can be implemented within existing gradient damage-based finite element frameworks with minimal implementation changes. The results highlight the potential of this approach to resolve the decades-long challenge of spurious widening in gradient damage models, offering an effective and practical solution for engineering applications.

\end{abstract}

\end{frontmatter}

\section{Introduction}
\label{Sec:Introduction}

\subsection{Literature Review}
\label{Sec:IntroLitReview}
Founded on the seminal work of Griffith's theory \cite{griffith1921vi}, discrete crack models constitute a compelling framework for simulating fracture, as they enable the explicit representation of cracks as geometric discontinuities within the material and reflect closely the real-world behavior of fracture initiation and propagation \cite{ngo1967finite,de2022fracture}. However, these benefits come at the expense of increased computational complexity and the associated costs of explicit crack tracking and remeshing during crack propagation \cite{stanic2020fracture, moes1999finite}. Though these limitations are partially addressed by advancements such as the Extended Finite Element Method (XFEM) \cite{belytschko1999elastic, wells2001new, moes1999finite}, XFEM encounters significant challenges when applied to three-dimensional problems, where accurately tracing complex crack paths remains difficult \cite{duarte2017numerical,gupta2017accuracy}. Furthermore, discrete fracture mechanics approaches face significant challenges when extended to coupled physics problems \cite{chen2022thermodynamically} such as poroelasticity or viscoplasticity due to the need for a fracture initiation criteria \cite{feng2019xfem} and non-traditional integration solution schemes \cite{wang2017direct}. These limitations motivated the emergence of smeared damage models as an alternative approach to fracture modeling, which became more popular since the 1980s \cite{rashid1968ultimate,jirasek2011damage,yoshioka2019comparative}. Within this family of methods, fracture is represented as a degradation of material stiffness rather than an explicit crack surface, and it is expressed by means of a continuous variable. The two most popular smeared damage approaches are phase-field \cite{miehe2010thermodynamically,kuhn2010continuum,bourdin2008variational} and non-local damage models \cite{peerlings1996gradient,pham2011gradient,peerlings1996some}, which are discussed below.

Phase-field fracture models emerged in the early 2000s as a regularized variational reformulation of Griffith's theory, aiming to capture all stages of crack evolution within a continuum domain. Originally developed by Bourdin, Francfort, and Marigo \cite{francfort1998revisiting,bourdin2000numerical,bourdin2008variational} and later extended by Miehe et al. \cite{miehe2010phase,miehe2010thermodynamically}, these models introduce a scalar phase-field variable ranging from 0 (representing intact material) to 1 (representing the fully damaged state). Despite the widespread adoption in recent years for various applications \cite{wick2017modified,peng2020phase,zhou2018phase,kristensen2020phase,de2022nucleation}, the phase-field approach faces two key challenges: a) the need for substantially fine meshes to resolve the diffusive crack zone, which leads to excessive computational costs \cite{heister2015primal,patil2018adaptive,geelen2020extended,freddi2022mesh}, and b) the need for mathematical reformulation when extended to new classes of material responses or to multi-physics applications \cite{miehe2016phase,aldakheel2021multilevel,kumar2020revisiting,lopez2025classical}.  

Integral-type nonlocal damage models \cite{bazant1984continuum,bazant1988nonlocal,de1995comparison}, commonly referred to as nonlocal damage models, were developed to overcome the limitations of local damage formulations, such as loss of ellipticity and pathological mesh sensitivity in the softening regime \cite{pijaudier1987nonlocal}. In these models, damage at a material point is computed as a weighted spatial average over an interaction zone that is defined by the material's characteristic length. However, the integral-type damage formulation introduces spurious boundary effects due to the truncation of the interaction zone near domain edges, often resulting in artificial stiffness or delayed damage initiation \cite{peerlings1996some}. Gradient-enhanced damage models were introduced  \cite{mobasher2018thermodynamic,mobasher2021dual,peerlings2004thermodynamically, peerlings1996gradient,desmorat2007nonlocal,poh2009over,geers2000damage} to address these limitations by introducing a Helmholtz type PDE that governed the non-local damage evolution, dependent on the characteristic length of the material, and restored the well-posedness of the boundary value problem. 

These formulations, while useful in overcoming the challenges of previous non-local models, led to spurious damage growth and widening in the damage zone. This was initially attributed to the fixed length scale used in these models, and several solutions were proposed to address the same. Localizing gradient damage models \cite{poh2017localizing, zhang2022simple, negi2022continuous} employed a damage- or strain-dependent length scale that progressively vanished in severely damaged regions to promote sharper localization. In contrast, transient length scale models \cite{geers1998strain, saroukhani2013simplified} introduced an evolving internal length scale that grew from zero or a minimum value to a finite maximum based on the equivalent strain, thereby controlling the extent of nonlocal interactions during damage evolution. Alternative strategies, such as employing multiple length scales to capture distinct failure modes \cite{ahmed2021nonlocal} and energy-limiter approaches \cite{tran2023new} that define the non-local damage and characteristic length scale based on physically motivated strain energy density, have also been proposed. Still other approaches have advocated for a hybrid strategy that transitions from a gradient damage formulation to a discrete cohesive model to mitigate the damage band widening issue \cite{Comi2007}. While effective at mitigating spurious damage artifacts and successful in specific applications, these methods serve as partial remedies and do not fully resolve the underlying cause of the widening damage band problem. Therefore, both from a theoretical and an implementation standpoint, a fundamental and comprehensive solution to this problem still remains elusive. 

De Borst et al. \cite{de2016gradient} proposed an explanation for the widening damage band phenomenon by contrasting gradient-enhanced non-local damage models with phase-field formulations. They demonstrated that, unlike phase-field models where the thermodynamic damage driving force diminishes to zero at full damage — thus leading to a fixed-width damage band — gradient models exhibit a non-vanishing thermodynamic damage driving force at full damage, thus resulting in a non-physical diffusion of damage beyond the process zone. Another hypothesis proposed by Duda et al. \cite{Duda2015} suggests that the inability of gradient damage models to accurately capture strong discontinuities arises from a rate driving force that fails to decay as damage progresses. However, this hypothesis was not examined in detail.

\subsection{Scope and Outline}
\label{Sec:IntroScopeOutline}

In this work, we present a new theoretical and thermodynamics-consistent framework that directly tackles the long-standing issue of the widening damage band in gradient models. We identify two fundamental causes for this issue: (a) the thermodynamic damage driving force being non-zero when damage reaches its maximum value, and (b) the absence of a decay function that attenuates non-local strain growth at peak damage. We demonstrate that these two features work together to cause continuous damage spread, and one must account for both to achieve constant, non-spreading damage bands. To illustrate this behavior, we investigate a series of approaches that tackle each aspect separately, and finally, we propose a novel modified non-local damage formulation that directly addresses both root causes. We emphasize that our implementation requires only minimal modifications to the conventional gradient damage framework, and it can therefore be readily implemented in existing solvers. 

The paper is structured as follows: Section \ref{Sec:Theoretical_background} outlines the theoretical background of the work. Section \ref{Sec:Methodology}, details the methodology of proposed model. In Section \ref{Sec:Numerical_Solvers}, the details of the FEM routine, constitutive modeling approaches and damage models used in this work are presented. Section \ref{Sec:NumericalExamples} illustrates the performance of this modified gradient damage model using several benchmark problems. Finally, Section \ref{Sec:Conclusion} presents a critical analysis of the results along with a summary of the key contributions and implications of this work. 


\section{Theoretical background: gradient damage vs phase-field models}
\label{Sec:Theoretical_background}


Non-local gradient and phase-field damage models both emerged as remedies for the mesh sensitivity and computational challenges faced by local damage and linear elastic fracture mechanics formulations, respectively. Although both approaches employ regularization strategies to overcome the ill-posedness of their predecessor models, their foundational principles and theoretical formulations differ significantly. Gradient damage models originate within the framework of classical continuum mechanics and seek to restore the well-posedness of boundary value problems in the post-peak softening regime by introducing non-local damage through a diffusion-type PDE that depends on the length scale \cite{peerlings1996gradient}. Phase-field fracture models emerged from the variational regularization of Griffith’s brittle‐fracture criterion: Ambrosio and Tortorelli \cite{ambrosio1990approximation} proposed approximating sharp crack discontinuities by a diffusive field within the Mumford–Shah functional, and Francfort and Marigo \cite{francfort1998revisiting} later recast brittle fracture as a global energy‐minimization problem. When applied to solid mechanics, this Ambrosio–Tortorelli (AT) framework yields a coupled system in which cracks evolve as smeared damage bands governed by a length scale. Although these two approaches stem from distinct motivations — mechanical non-locality versus mathematical regularization — they converge on analogous partial differential equation formulations that capture damage localization, as demonstrated in the work of de Borst \cite{de2016gradient}. 

Conventional gradient damage ({\bf{CGD}}) models \cite{peerlings1996gradient} offer two notable advantages over phase-field models. First, CGD models can effectively regularize damage even with relatively coarse meshes, typically requiring a length scale ($l_c$) to element size ($h)$ ratio of $l_c/h \approx 2.0$ \cite{Mandal2019}. By contrast, phase-field models demand much finer discretization, with $l_c/h \approx 6.0$ \cite{kristensen2021assessment,wu2017unified}, to accurately capture the dissipation of fracture energy near the crack tip. This ability to operate with coarser meshes makes CGD models appealing for simulations where fine discretization can be computationally prohibitive. Secondly, CGD models can directly incorporate empirical damage evolution laws such as those by Mazars \cite{mazars1986description}, Geers \cite{geers1998strain},  Ahmed et al. \cite{ahmed2021nonlocal,Ahmed2020DamagedPlasticity} and Ortiz et al. \cite{ortiz1985constitutive}, some of which have been widely adopted and validated against experimental data over the past decades. This allows for straightforward integration of newly developed damage laws tailored to specific materials or loading scenarios, with minimal to no changes in the CGD framework, rendering this approach highly adaptable and suitable for a broad range of applications. By contrast, phase-field models do not prescribe damage evolution directly; it arises as a consequence of the chosen form of the Helmholtz free energy functional. As a result, incorporating new material behavior requires a complete reformulation of the energy functional to accurately capture the underlying physics of the material. This process is often non-trivial and demands significant theoretical understanding and computational effort. The complexity of this task is evident in the diverse phase-field formulation attempts proposed to describe similar material responses as discussed in \cite{kumar2020revisiting,lopez2025classical,alessi2018comparison} and the references therein.

\begin{table}[H]
\renewcommand{\arraystretch}{1.6}
\centering
\caption{Comparison of the degradation function $(g)$, thermodynamic damage driving force ($Y$) and forcing term in the CGD and phase-field model}
\begin{tabularx}{\textwidth}{*{4}{>{\centering\arraybackslash}X}}
\hline \hline
\textbf{Model} & \textbf{Degradation function} $g$ & \textbf{Thermodynamic damage driving force} $Y$ & \textbf{Forcing term}  \\ 
\hline
Conventional Gradient Damage (CGD) & $g(d) = 1 - d$ & $\dfrac{1}{2}\varepsilon:\boldsymbol{C}:\varepsilon$ & $\varepsilon$\\
\hline
Phase-field  & $g(\phi) = (1-\phi)^2$ & $\dfrac{1}{2}g'(\phi)\varepsilon:\boldsymbol{C}:\varepsilon$ & $ g'(\phi)\varepsilon:C:\varepsilon$ \\
\hline
\end{tabularx}
\label{Table:CGDvsPhase}
\end{table}

A comparison of both methods' formulation, particularly in terms of the degradation function, thermodynamic damage driving force and forcing term, is presented in Table \ref{Table:CGDvsPhase}. An explanation of the symbols follows: 

\begin{itemize}
    
    \item $d$ and $\phi$ are the damage and phase-field variables respectively, both ranging between 0 (intact material) and 1 (cracked state)

    \item $\boldsymbol{C}$ is the fourth-order elasticity tensor (see Eqn. \eqref{Elasticity_tensor_defination})

    \item $\varepsilon$ is the local strain tensor, for which several models are available (see Eqns. \eqref{Mazars_Equivalent_strain_tension} and \eqref{deVree_strain_definition}) 

    \item $Y$ is the thermodynamic damage driving force, which is the derivative of the Helmholtz free energy with respect to the evolving damage variable
    
\end{itemize}

Here we observe that the forcing term in both approaches depends on the local strain $\varepsilon$, but in the phase-field model the forcing term and the thermodynamic damage driving force are also functions of the phase-field variable $\phi$. This characteristic is not present in the CGD method, where $Y$ is independent of $d$. The authors in \cite{de2016gradient} proposed that the widening damage band issue observed in CGD models is due to the non-zero thermodynamic damage driving force when damage approaches the maximum value of 1; however, the study did not test this hypothesis.


\section{Methodology}
\label{Sec:Methodology}

In this section, we present the initial strategies investigated to address the widening damage band issue inherent to conventional gradient damage (CGD) models, guided by observations from the literature. We then introduce our final formulation, which offers a thermodynamically consistent solution and demonstrates superior performance compared to existing approaches.

\subsection{Initial solution approaches}

The first approach explored in this study, referred to as {\bf{modA}}, aims to address the issue identified by de Borst \cite{de2016gradient}—namely, the presence of a non-zero thermodynamic damage driving force when $d \approx 1$. To resolve this, the standard degradation function applied to the stress definition is modified by introducing an auxiliary function that depends on the damage variable at the current and previous load steps $k$, denoted as $d$ and $\prescript{k-1}{}{d}$ respectively. A sample expression for this function reads:

\begin{equation}
    f_a = f(d,\prescript{k-1}{}{d})=\dfrac{d^2}{2} - \dfrac{{\prescript{k-1}{}{d}}^2}{2}
\end{equation}

This auxiliary function $f_a$ is designed to have a minimal influence on the stress response while ensuring a phase-field like decay in the thermodynamic damage driving force $Y$. The corresponding Helmholtz free energy expression and the resulting form of $Y$ are presented in Tables \ref{Tab:Helmholtz_comparison} and \ref{Tab:MNLDvsinitial} respectively, where we note that $h$ is a stiffness variable that controls the non-local damage energy storage \cite{peerlings2004thermodynamically}. Additionally, Table \ref{Tab:MNLDvsinitial} summarizes the degradation functions and forcing terms employed across all models. A detailed investigation of the performance of the {\bf{modA}} model is presented in Section \ref{Sec:NumericalExamples}. Despite the decay of $Y$ to zero in the modA model, our results show that the overall response of modA does not differ appreciably from that of the CGD in the 1D example (Section \ref{NumEx: 1D problems}), while facing convergence issues in the 2D case (Section \ref{SNS_compare_modA_modB}). As discussed in more detail in these sections, these observations suggest that the widening of the damage band in CGD models cannot be attributed solely to the non-zero thermodynamic damage driving force at complete damage. Accordingly, our subsequent investigations were directed toward identifying additional contributing factors responsible for this pathological behavior.

Our next approach, named {\bf{modB}}, was inspired by the phase-field model that has a forcing term that decays with damage evolution. In modB, we modified the CGD framework by introducing a decay function $f_r$ that affects the forcing term $\varepsilon$ such that it remains unaffected over most of the damage evolution regime, but smoothly decays to zero as damage $d$ approaches 1. The expression for this function reads: 

\begin{equation}
    f_r = 1 - d^n
\end{equation}

\begin{figure}[b!]
\centering
\includegraphics[width=1.1\textwidth,trim = 2cm 4cm 2cm 4cm, clip]{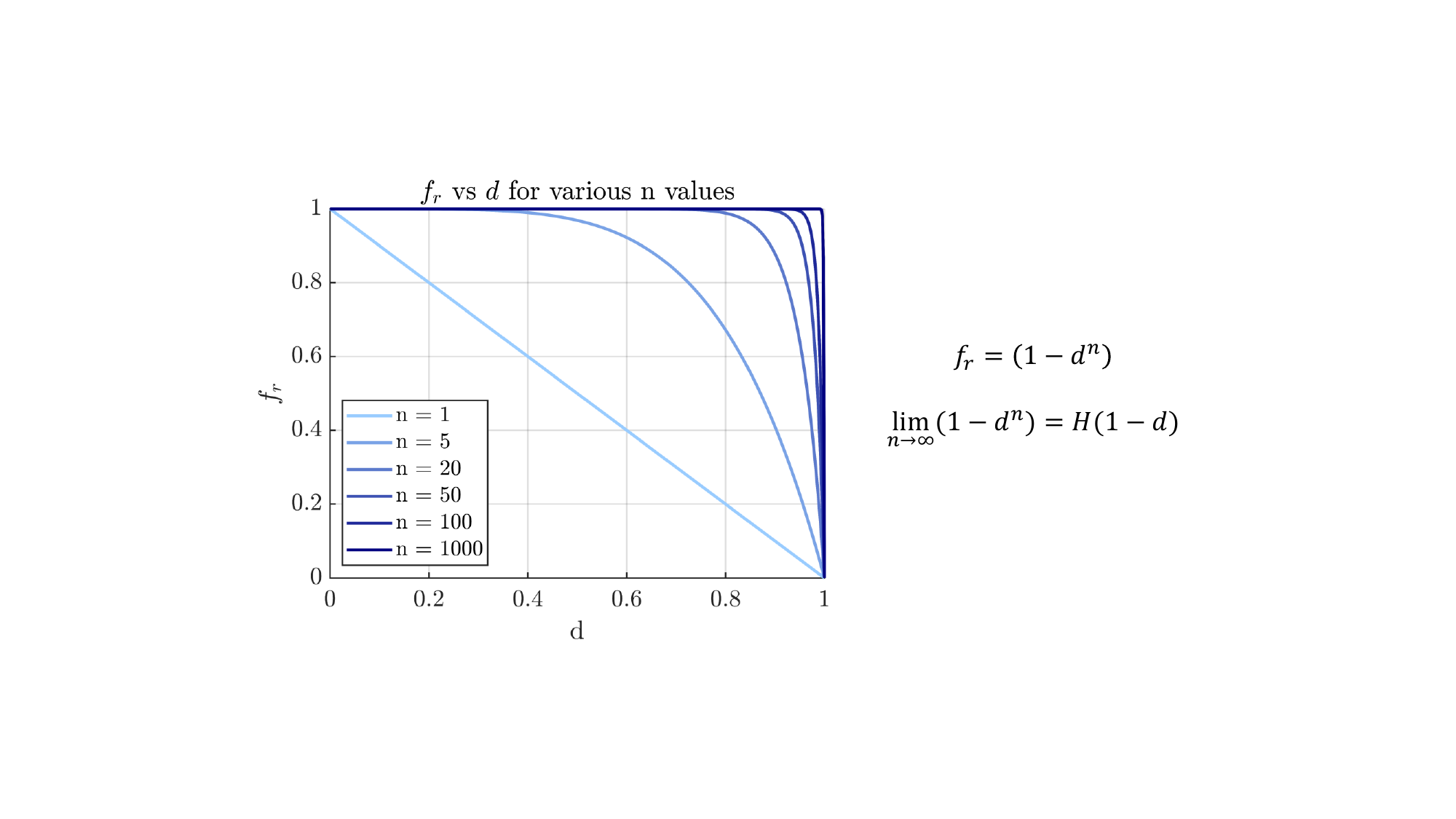}
\caption{Schematic representation of the decay function $f_r$ applied to the local strain $\varepsilon$ vs damage. As damage approaches the maximum value of 1 for an $n$ approaching infinity, $f_r$ approaches the behavior of the Heaviside function $H(x)$ where $x=(1-d)$.}
\label{fig:fr_crossplot}
\end{figure}

Fig. \ref{fig:fr_crossplot} shows the evolution of $f_r$ for different values of the user-defined exponent $n$. As $n$ increases, the behavior of $f_r$ asymptotically approaches that of the Heaviside function $H(1-d)$, thus influencing the forcing term only near the terminal value of damage. In Section \ref{Sec:NumericalExamples}, the numerical examples previously used to evaluate modA are also employed to assess the performance of modB. Notably, while modB performs similarly to modA and CGD in the one-dimensional example (Section \ref{NumEx: 1D problems}), it successfully limits the spread of the damage zone in the two-dimensional problem presented in Section \ref{SNS_compare_modA_modB}. However, despite this improvement, modB suffers from the same thermodynamic inconsistency as the CGD model, as the driving force $Y$ remains unaffected by damage evolution.

\begin{table}[t!]
\renewcommand{\arraystretch}{2}
\centering
\caption{Comparison of the Helmholtz free energy expression in the CGD, modA, modB and MNLD models.}
\begin{tabular}{c c}
\hline \hline 
\textbf{Model} & \vspace{-0.75cm} \textbf{Helmholtz free energy expression}  \\
 & $\psi (\varepsilon_{ij},\bar\varepsilon,d) $ \\
\hline
Conventional Gradient Damage (CGD) & $\dfrac{1}{2} (1-d) \varepsilon_{ij} C_{ijkl} \varepsilon_{kl} + \dfrac{1}{2}h \left[({\varepsilon} - \bar\varepsilon)^2 + c \bar\varepsilon_{,i}^2 \right]$\\
\hline
modA & $\dfrac{1}{2} \left[(1-d) + \boldsymbol{f(d,\prescript{k-1}{}{d}))} \right] \varepsilon_{ij} C_{ijkl} \varepsilon_{kl} + \dfrac{1}{2}h \left[({\varepsilon} - \bar\varepsilon)^2 + c \bar\varepsilon_{,i}^2 \right]$\\
\hline
modB & $\dfrac{1}{2} (1-d) \varepsilon_{ij} C_{ijkl} \varepsilon_{kl} + \dfrac{1}{2}h \left[({  \boldsymbol{(1-d^{n})} \varepsilon} - \bar\varepsilon)^2 + c \bar\varepsilon_{,i}^2 \right]$\\
\hline
Modified Non-Local Damage (MNLD) & $\dfrac{1}{2} [(1-d) + \boldsymbol{f(d,\prescript{k-1}{}{d})}] \varepsilon_{ij} C_{ijkl} \varepsilon_{kl} + \dfrac{1}{2}h \left[{( \boldsymbol{(1-d^{n})} \varepsilon} - \bar\varepsilon)^2 + c \bar\varepsilon_{,i}^2 \right]$\\
\hline
\end{tabular}
\label{Tab:Helmholtz_comparison}
\end{table}

\begin{table}[H]
\renewcommand{\arraystretch}{2}
\centering
\caption{Comparison of the degradation function $g$, thermodynamic force ($Y$) and forcing term in the modA, modB and MNLD models. Here, $f_a = f(d,\prescript{k-1}{}{d})=\dfrac{d^2}{2} - \dfrac{{\prescript{k-1}{}{d}}^2}{2}$ and $f_r = 1 - d^n$, where $n$ is a user-defined exponent. In our work the stiffness parameter $h$ is set to zero.}
\begin{tabular}{c  c  c  c }
\hline \hline 
\textbf{Model} & \textbf{Degradation function $g$} & ${Y} = - \dfrac{\partial \psi}{\partial d}$ & \textbf{Forcing term} \\ 
\hline 
CGD & $(1-d)$ & $ \dfrac{1}{2}\varepsilon:\boldsymbol{C}:\varepsilon$ & $\varepsilon$\\
\hline
modA & $(1-d) + f_a$ & $  \left(1-\dfrac{\partial f(d,\prescript{k-1}{}{d})}{\partial d} \right)\dfrac{1}{2}\varepsilon:\boldsymbol{C}:\varepsilon$ & $\varepsilon$\\
\hline
modB & $(1-d)$ & $ \dfrac{1}{2}\varepsilon:\boldsymbol{C}:\varepsilon$ &  $f_r   \varepsilon$\\
\hline
MNLD & $(1-d) + f_a$ & $ \left(1-\dfrac{\partial f(d,\prescript{k-1}{}{d})}{\partial d} \right) \dfrac{1}{2}\varepsilon:\boldsymbol{C}:\varepsilon$ & $f_r \varepsilon$\\
\hline
\end{tabular}
\label{Tab:MNLDvsinitial}
\end{table}

\subsection{Adopted approach}

Building upon the insights gained from modA and modB, we hypothesize that a thermodynamically consistent solution to the widening damage band problem observed in CGD can be achieved if the following conditions are satisfied:

\begin{itemize}
    \item The thermodynamic damage driving force $Y$ vanishes as the damage reaches $1$.
    \item The evolution of the forcing term $\varepsilon$ is attenuated as damage approaches 1.
\end{itemize}

\begin{figure}[t!]
\centering
\includegraphics[width=1\textwidth,trim = 3.5cm 7cm 3.5cm 7cm, clip]{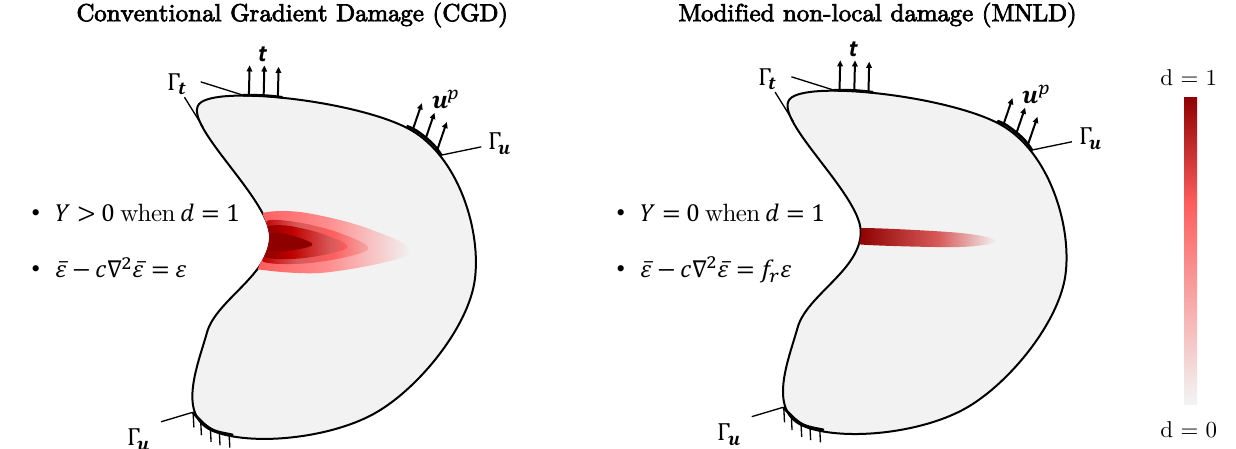}
\caption{Schematic representation of the conventional gradient damage (CGD) model and modified non-local damage (MNLD).}
\label{CGDvsMNLD_schematic}      
\end{figure}

To satisfy the conditions stated above, we propose a combined formulation involving two key modifications to the Helmholtz free energy expression of the CGD model: a) the introduction of an auxiliary function $f(d,\prescript{k-1}{}{d})$ in the stress degradation function definition, and b) the addition of a decay function applied to the forcing term $\varepsilon$. This framework, which combines aspects of modA and modB, is referred to as the modified non-local damage ({\bf{MNLD}}) model. A schematic comparison of the CGD and MNLD formulations, with idealized damage profiles, is provided in Fig. \ref{CGDvsMNLD_schematic}.  

The performance of the MNLD model, in comparison to CGD, modA, and modB, is assessed in detail in Section \ref{Sec:NumericalExamples}. The MNLD model addresses a fundamental inconsistency that has persisted in the gradient-enhanced non-local damage literature, marking a significant step forward in smeared damage modeling. We also emphasize that the MNLD framework requires only minimal modifications to the conventional approach, which makes its implementation straightforward within existing continuum damage mechanics codes. The derivation and FEM framework of the adopted model are presented in the next section.


\section{FEM formulation}
\label{Sec:Numerical_Solvers}

In this section we outline the governing equations and constitutive models employed in this study. The numerical implementation is carried out using a monolithic Newton-Raphson solution scheme. The full derivation from the Helmholtz free energy expression to the governing equations is presented in \ref{Apx:psi_defination}. 

\subsection{Governing equations}

\subsubsection{Strong form expressions} 

Consider a schematic domain $\Omega$, as shown in Fig.  \ref{CGDvsMNLD_schematic}, bounded by $\Gamma = \Gamma_u \cup \Gamma_t$, where $\Gamma_u$ represents the portion of the boundary subjected to prescribed displacements $\boldsymbol{u}^p$, and $\Gamma_t$ denotes the segment where external traction $\boldsymbol{t}$ is imposed. Assuming small deformations and using the Einstein notation, the strain and elasticity tensor are defined as: 

\begin{equation} 
 \varepsilon_{ij}= \frac{1}{2} \left[u_{i,j} + u_{j,i} \right]
\label{Strain_tensor_defination}
\end{equation}

\begin{equation}
 C_{ijkl} = \left[K \ {\delta}_{ij} {\delta}_{kl} + \mu \left[ {\delta}_{ik} {\delta}_{jl} + {\delta}_{il} {\delta}_{jk} - \frac{2}{3} {\delta}_{ij} {\delta}_{kl} \right] \right] ; 
\label{Elasticity_tensor_defination}
\end{equation}

\noindent where $K$ is the bulk modulus, $\mu$ is the shear modulus, $\delta_{ij}$ is the Kronecker delta function, $C_{ijkl}$ is the fourth-order elasticity tensor, $\varepsilon_{ij}$ is the local strain tensor. 

As discussed in Section \ref{Sec:Methodology}, in this work we introduce two new functions compared to the conventional gradient damage approach: 

\begin{itemize}

    \item $f_a = f(d,\prescript{k-1}{}{d})=\dfrac{d^2}{2} - \dfrac{{\prescript{k-1}{}{d}}^2}{2}$: a function which complements the nominal stress degradation function to ensure thermodynamic consistency

    \item $f_r = 1 - d^{n}$: a function which is applied to the forcing term $\varepsilon$ to attenuate it progressively as damage evolves to its peak value

\end{itemize}

As a result of the first feature, the modified stress ${\sigma}_{ij}$ is now expressed as: 

\begin{equation}
\sigma_{ij}  
 = 
\left[(1-d) + f(d,\prescript{k-1}{}{d}) \right] C_{ijkl} \varepsilon_{kl} 
\label{Equation 4}
\end{equation}

\noindent where, ${\sigma}_{ij}$ is the Cauchy stress tensor. Here, damage $d=d(\bar \varepsilon)$ is a function of the non-local strain. Following the above, the strong form of the balance of momentum and strain-diffusion differential equation can then be expressed as: 

\begin{equation}
    {\sigma}_{ij,j}=0
\label{CDM_Strong_Form_1}
\end{equation}

\begin{equation}
    \bar{\varepsilon} - c {{\bar {\epsilon}},}_{ii} = f_r \varepsilon
\label{CDM_Strong_Form_2}
\end{equation}

\noindent where $\bar{{\varepsilon}}$ is the non-local strain, ${{\bar \varepsilon},}_{ii}$ is the second-order spatial partial derivative of the non-local strain, and $c$ is a length scale variable which is a function of the characteristic length $l_{c}$ ($c = {l_{c}}^{2}/2$). The function $f_r$ serves as a decay term that progressively attenuates the forcing term $\varepsilon$ as damage approaches its peak value, and $n$ represents the exponent controlling the rate of decay.

\subsubsection{Discretized weak forms}

Following standard FEM procedures, an  arbitrary weight function $\boldsymbol{w}^u$, shape function ${\boldsymbol{N}}^u$ and its derivative ${\boldsymbol{B}}^u$ are introduced to formulate the discretized weak form from the strong form Eqn. \eqref{CDM_Strong_Form_1}:

\begin{equation}
\boldsymbol{u} = \boldsymbol{N}^u u^e  \qquad ;  \qquad \boldsymbol{w}^u =  \boldsymbol{N}^u w^{u,e} \qquad ; \qquad \nabla \boldsymbol{w}^u = \boldsymbol{B}^u w^{u,e}
\label{Disp_shape_functions}
\end{equation}

\noindent where the superscript 'e' refers to nodal values. The nodal discretization of the non-local strain requires the introduction of weight functions, shape functions, and their derivatives to Eqn. \eqref{CDM_Strong_Form_2} as follows; details of the natural boundary condition on Eqn. \eqref{CDM_Strong_Form_2} are available in \cite{peerlings1996gradient}:

\begin{equation}
{\boldsymbol{\bar \varepsilon}} = \boldsymbol{N}^{\bar \varepsilon} {{\bar \varepsilon}}^{e} \qquad ; \qquad \boldsymbol{w}^{\bar \varepsilon} = \boldsymbol{N}^{\bar \varepsilon} w^{{\bar \varepsilon},e} \qquad ; \qquad \nabla \boldsymbol{w}^{\bar \varepsilon} = \boldsymbol{B}^{\bar \varepsilon} w^{{\bar \varepsilon},e}
\label{Nonlocal_shape_functions}
\end{equation} 

Applying Eqns. \eqref{Disp_shape_functions} and \eqref{Nonlocal_shape_functions} to the governing Eqns. \eqref{CDM_Strong_Form_1} and \eqref{CDM_Strong_Form_2} yields the following discretized weak form of the governing PDEs: 

\begin{equation}
\boldsymbol{r}^u = \underbrace{\int_\Omega [\boldsymbol{B}^u]^T \boldsymbol{\sigma} d\Omega}_{\boldsymbol{f}^{int}} - \underbrace{\int_\Gamma [\boldsymbol{N}^u]^T t d\Gamma}_{\boldsymbol{f}^{ext}} 
\label{r_u_Nonlocal_weak}
\end{equation}

\begin{equation}
\boldsymbol{r}^{\bar \epsilon} = \underbrace{{\int_\Omega [\boldsymbol{N}^{\bar \epsilon}]^T \boldsymbol{\bar{\epsilon}} \ d\Omega + \int_\Omega [\boldsymbol{B}^{\bar \epsilon}]^T \ c \ \nabla \boldsymbol{\bar{\epsilon}} \ d\Omega} -{\int_\Omega [\boldsymbol{N}^{\bar \epsilon}]^T \boldsymbol{\epsilon} \ d\Omega}}_{\boldsymbol{f}^{int, \bar\epsilon}} 
\label{r_e_Nonlocal_weak}
\end{equation}

Minimizing the residual $\boldsymbol{r}^u$ and $\boldsymbol{r}^{\bar \varepsilon}$ in Eqn. \eqref{r_u_Nonlocal_weak} and \eqref{r_e_Nonlocal_weak} leads to the solution of the non-linear system. We note that the terms $\boldsymbol{f}^{int}$ and $\boldsymbol{f}^{ext}$ refer to the internal and external force vectors respectively. The term $\boldsymbol{f}^{int, \bar \varepsilon}$ refers to the internal force vector due to the non-local strain. The overall system of equations to be minimized can be expressed as: 

\begin{equation}
\boldsymbol{r}= \boldsymbol{f}^{int}-\boldsymbol{f}^{ext} = - \boldsymbol{J} \delta \boldsymbol{x}
\label{General_residual_vector}
\end{equation}

\noindent where $\boldsymbol{J}$ is the Jacobian matrix of the system defined as $\boldsymbol{J}={\partial\boldsymbol{r}}/{\partial\boldsymbol{x}}$ and $ \{\boldsymbol{x}\} = [\boldsymbol{u} \quad \boldsymbol{\bar \varepsilon}]^T$. Here, it is important to note that damage evolution — which directly influences the stress tensor $\boldsymbol \sigma$ - proceeds in a direction consistent with the enforcement of the Clausius-Duhem inequality. This inequality embodies the first and second laws of thermodynamics, ensuring the irreversibility of damage growth \cite{murakami2012continuum,kachanov1986introduction}. In continuum damage mechanics formulations, this requirement is enforced by imposing the following condition, where $\dot d$ refers to the damage growth rate:

\begin{equation}
\dot d \geq 0 
\label{Kuhn-Tucker_conditions}
\end{equation}

When using the Newton-Raphson solver to solve Eqn. \eqref{General_residual_vector}, the following system of equations emerge:
\begin{equation}
\begin{bmatrix}
\boldsymbol{J}_{uu}                  & \boldsymbol{J}_{u \bar{\varepsilon}} \\
\boldsymbol{J}_{\bar{\varepsilon} u} & \boldsymbol{J}_{\bar{\varepsilon} \bar{\varepsilon}}
\end{bmatrix}
\begin{bmatrix}
\delta \boldsymbol{u} \\
\delta \boldsymbol{\bar{\varepsilon}}
\end{bmatrix}
=
-
\begin{bmatrix}
\boldsymbol{r}^u \\
\boldsymbol{r}^{\varepsilon}
\end{bmatrix}
\end{equation}

\noindent where $\delta \boldsymbol{u}$  and $\delta \boldsymbol{\bar{\varepsilon}}$ are the incremental nodal displacements and non-local equivalent strains, and the entries of the Jacobian matrix are:

\begin{align}
\quad \boldsymbol{J}_{uu} &= \int_\Omega (\boldsymbol{B}^u)^T \frac{\partial \boldsymbol{\sigma}}{\partial u^e} d\Omega = \int_\Omega (\boldsymbol{B}^u)^T \left[(1 - d) + f(d,\prescript{k-1}{}{d})\right] C_{ijkl} \boldsymbol{B}^u  d\Omega
\end{align}

\begin{align}
\quad \boldsymbol{J}_{u \bar{\epsilon}} &=
\int_\Omega (\boldsymbol{B}^u)^T \dfrac{\partial \boldsymbol{\sigma}}{\partial \bar{\epsilon}^e} 
d\Omega = \int_\Omega (\boldsymbol{B}^u)^T \left[
- 
 \dfrac{\partial d} {\partial \bar\varepsilon} 
 +
\dfrac{\partial f(d,\prescript{k-1}{}{d})} {\partial d} 
 \dfrac{\partial d} {\partial \bar\varepsilon} 
 \right]  C_{ijkl} \epsilon_{kl} \boldsymbol{N}^{\bar\varepsilon} d\Omega
\label{eq_Jue}
\end{align}

\begin{align}
\quad \boldsymbol{J}_{\bar{\varepsilon} u} &= - \int_\Omega (\boldsymbol{N}^{\varepsilon})^T f_r \dfrac{\partial \varepsilon}{\partial \varepsilon_{ij}} \boldsymbol{B}^u
 \ d\Omega 
\end{align}

\begin{align}
\quad \boldsymbol{J}_{\bar{\varepsilon} \bar{\varepsilon}} & =
\int_\Omega (\boldsymbol{N}^{\varepsilon})^T (\boldsymbol{N}^{\varepsilon}) 
- 
\int_{\Omega} (\boldsymbol{N}^{\bar\varepsilon})^T \varepsilon \dfrac{\partial f_r}{\partial d} \dfrac{\partial d}{\partial \bar\varepsilon} \dfrac{\partial \bar\varepsilon}{\partial \bar\varepsilon^e}
+  
(\boldsymbol{B}^{\varepsilon})^T c\boldsymbol{B}^{\varepsilon} d\Omega 
\end{align}

\subsection{Constitutive modeling}

In this subsection, we briefly describe the damage modeling strategies adopted in this study, which include two distinct damage evolution laws and two non-local strain definitions. These approaches are integrated within the finite element method (FEM) framework.

\subsubsection{Damage laws}
\label{Sec:Damage_laws}

Exponential damage laws, like those proposed by Mazar \cite{mazars1986description} and Geers et al. \cite{geers1998strain}, asymptotically approach the maximum value of 1 without actually attaining it within the loading scenario of most problems. A modification to these damage laws is adopted here to enforce $d=1$ based on two user-defined variables $s_1$ and $s_2$. Here, $s_1 = \dfrac{\varepsilon_f}{\varepsilon_{trans}}$ and $s_2 = \dfrac{\varepsilon_{trans}}{\varepsilon_{D}}$, where $\varepsilon_{trans}$ is a transition strain after which the evolution deviates from its conventional path and $\varepsilon_f$ is the strain at which damage reaches its full value of 1.

The definitions of the modifed Mazars and Geers damage laws are given in Eq. \eqref{modified_Mazars_damage} and Eqn. \eqref{modified_Geers_damage} respectively:

\begin{equation}
d(\bar\varepsilon) =
\begin{cases}
0, & \bar\varepsilon < \varepsilon_D \\[8pt]
1 - \dfrac{\varepsilon_D (1 - \alpha)}{\bar\varepsilon} - \dfrac{\alpha}{e^{\beta (\bar\varepsilon - \varepsilon_D)}}, & \varepsilon_D \leq \bar\varepsilon \leq \varepsilon_{d_{\text{trans}}} \\[8pt]
d_{\text{trans}} + (1 - d_{\text{trans}}) \left( \dfrac{\bar\varepsilon - \varepsilon_{d_{\text{trans}}}}{\varepsilon_f - \varepsilon_{d_{\text{trans}}}} \right)^{0.8}, & \varepsilon_{d_{\text{trans}}} < \bar\varepsilon \leq \varepsilon_f \\[8pt]
1, & \bar\varepsilon > \varepsilon_f
\end{cases}
\label{modified_Mazars_damage}
\end{equation}

\begin{equation}
d(\bar\varepsilon) =
\begin{cases}
0, & \bar\varepsilon < \varepsilon_D \\[8pt]
1 - \dfrac{\varepsilon_{D}}{\bar\varepsilon} \left\{ (1 - \alpha) + \alpha e^{\beta(\varepsilon_{D} - \bar\varepsilon)} \right\}, & \varepsilon_D \leq \bar\varepsilon \leq \varepsilon_{d_{\text{trans}}} \\[8pt]
d_{\text{trans}} + (1 - d_{\text{trans}}) \left( \dfrac{\bar\varepsilon - \varepsilon_{d_{\text{trans}}}}{\varepsilon_f - \varepsilon_{d_{\text{trans}}}} \right)^{0.8}, & \varepsilon_{d_{\text{trans}}} < \bar\varepsilon \leq \varepsilon_f \\[8pt]
1, & \bar\varepsilon > \varepsilon_f
\end{cases}
\label{modified_Geers_damage}
\end{equation}

\noindent where, $\varepsilon_{D}$ is the damage threshold strain and $\bar\varepsilon$ is a history variable (non-local strain). $\alpha$ and $\beta$ are damage model parameters that govern the response of the domain and are key variables in the calibration process along with $\varepsilon_D$ in the equations above. 

\begin{figure}[t!]
\centering
\includegraphics[width=1\textwidth,trim = 4cm 6.5cm 4cm 6.5cm, clip]{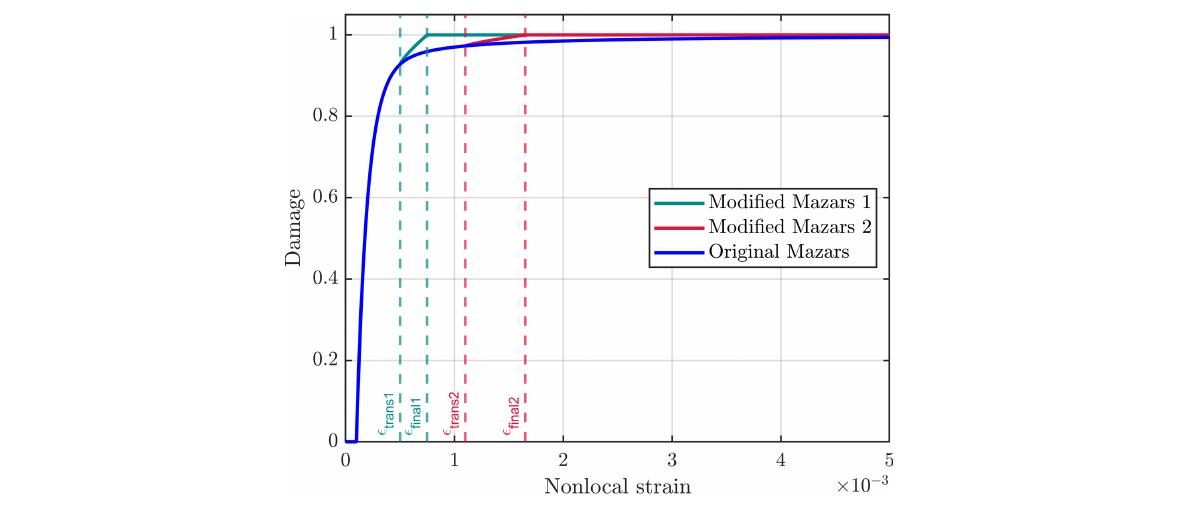}
\caption{Comparison of the classical Mazars damage law with the modified Mazars formulation proposed in this work. Modified Mazars models 1 and 2 use parameters s1 = \{1.5, 1.5\} and s2 = \{5,11\}, respectively.}
\label{fig:MazarsvsModifiedmazars}
\end{figure}

Fig. \ref{fig:MazarsvsModifiedmazars} illustrates the evolution of damage with respect to non-local strain, comparing the proposed modified Mazars damage law with the conventional formulation for two different combinations of $s_1$ and $s_2$. For a fixed value of $s_1$, increasing $s_2$ delays the onset of deviation from the conventional damage evolution, shifting the transition to a later stage in the damage path. A numerical example in Section \ref{subsec:SNS_s1_s2_study} details the impact of $s_1$ and $s_2$ on the performance of the proposed framework.

\subsubsection{Local equivalent strain}

Two variations of the local equivalent strain are implemented in this work. The first is calculated following Mazars approach \cite{mazars1984application}: 

\begin{equation}
  \varepsilon^{*}_{eq} = \sqrt{\sum_{P=1}^{3} {\langle \varepsilon_P \rangle}^2} ,
  \label{Mazars_Equivalent_strain_tension}
\end{equation}

\noindent where $\varepsilon_P$, $P$ = 1,2,3, are the principal strains, and the Macaulay brackets denote  $\displaystyle {\langle \boldsymbol{\cdot} \rangle} = \frac{\lvert \boldsymbol{\cdot} \rvert + \boldsymbol{\cdot} }{2}$. 

The second is the modified von Mises equivalent strain and its expression is given below \cite{de1995comparison}:
\begin{equation}
\varepsilon^{*}_{eq} = \dfrac{k_0-1}{2k_0(1 - 2\nu)} I_1^e + \dfrac{1}{2k_0} \sqrt{\left(\dfrac{k_0 - 1}{(1 - 2\nu)} I_1^e\right)^2 + \dfrac{12k_0}{(1 + \nu)^2} J_2^e}
\label{deVree_strain_definition}
\end{equation}

\noindent where, $k_0$ is a control parameter, $\nu$ is the Poisson's ratio, \( I_1^e \) is the first strain invariant and \( J_2^e \) is the second deviatoric invariant of strain. 


\section{Numerical Examples}
\label{Sec:NumericalExamples}

This section presents a numerical investigation of the modified non-local damage (MNLD) model and comparison against the conventional gradient damage (CGD) model under various scenarios. First, a one-dimensional bar under tension is analyzed as an initial verification benchmark. This example demonstrates the ability of the MNLD formulation to ensure that as $d$ approaches its maximum value, $Y$ drops to zero and the forcing term vanishes as well. Next, three 2D problems - Single Notch Shear (SNS), L-shaped panel, and three-point bending are presented to demonstrate the ability of the MNLD model to capture a fixed-width damage evolution, in contrast to the CGD model, which exhibits a progressively widening damage band during damage evolution. 

In each example, information regarding material properties, model parameters and modified damage evolution law are provided. Linear two-noded elements are used in 1D analyses, while bilinear four-noded elements are employed in all 2D examples. A quasi-static displacement-driven load and the monolithic Newton-Raphson solution scheme with adaptive load-stepping are used in all cases, with an absolute convergence tolerance of $10^{-6}$ being applied throughout. The loadfactor is defined as $\lambda = \dfrac{u}{u^p}$, where $u$ is the displacement at the current increment and $u^p$ is the total displacement load. All the analyses are conducted on a Dell Precision 7920 tower workstation with an Intel \textregistered Xeon \textregistered Gold 6230R CPU @ 2.1 GHz GHz, 64 GB RAM processor with 873.6 GFLOPS computing power \cite{intel_processors}.



\subsection{1D bar under tension}
\label{NumEx: 1D problems}

\begin{figure}[H]
    \centering
    \begin{minipage}{15cm}
    \centering     
    \includegraphics[width=1\textwidth,trim = 3.5cm 8cm 2cm 7cm, clip]{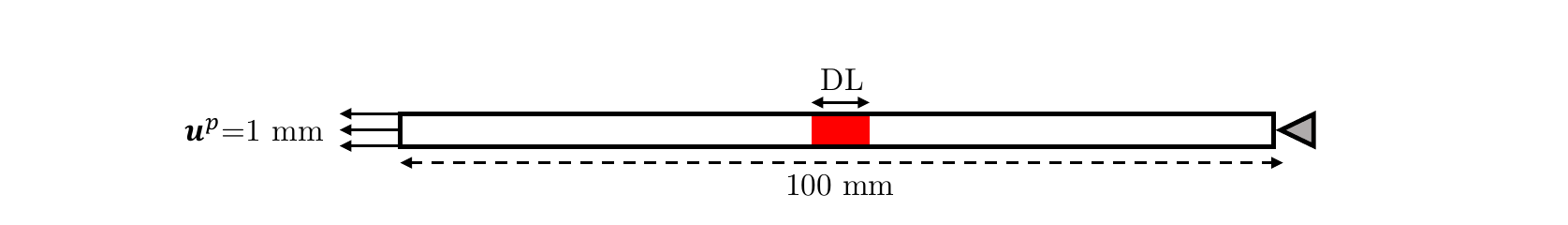}
    \end{minipage}
    \caption{Geometry and boundary conditions for the 1D bar problem. DL refers to the damaged length of the domain.}
    \label{fig:1D_bar_schematic}
\end{figure}

The first numerical example is a benchmark 1D bar under tension, aimed at verifying the implementation feasibility and examining the accuracy of the MNLD. Similar examples are reported in the relevant literature \cite{nguyen2018smoothing,poh2017localizing,saji2024new}. The bar, illustrated in Fig. \ref{fig:1D_bar_schematic}, measures 100 mm in length, has a unit cross-sectional area ($A = 1 mm^{2}$), is fixed at one end, and is subjected to a prescribed displacement of 1 mm at the opposite end. The FEM model comprises 201 elements, the elastic modulus is $E$ = 30 GPa, Poisson's ratio is $\nu=0.2$, and the modified Mazars damage model is adopted with $\alpha=0.7$, $\beta=10^{-4}$, $s_1=1.5$ and $s_2=5$. The Mazars equivalent strain definition (Eqn. \eqref{Mazars_Equivalent_strain_tension}) is used in this example. To initiate strain localization, a 4 mm damaged length (DL) is introduced at the center of the bar, such that $E_{DL} = 0.1 E$. The initial load step-size ($\Delta \lambda_0$) is $10^{-6}$ and the maximum load step-size ($\Delta \lambda_{max}$) is set to $10^{-4}$. Finally, the decay function exponent is chosen as $n = 10^{4}$. 

\begin{figure}[H]
    \centering
    \begin{minipage}{15cm}
    \centering     
    \includegraphics[width=1\textwidth,trim = 4.5cm 4cm 4.5cm 4cm, clip]{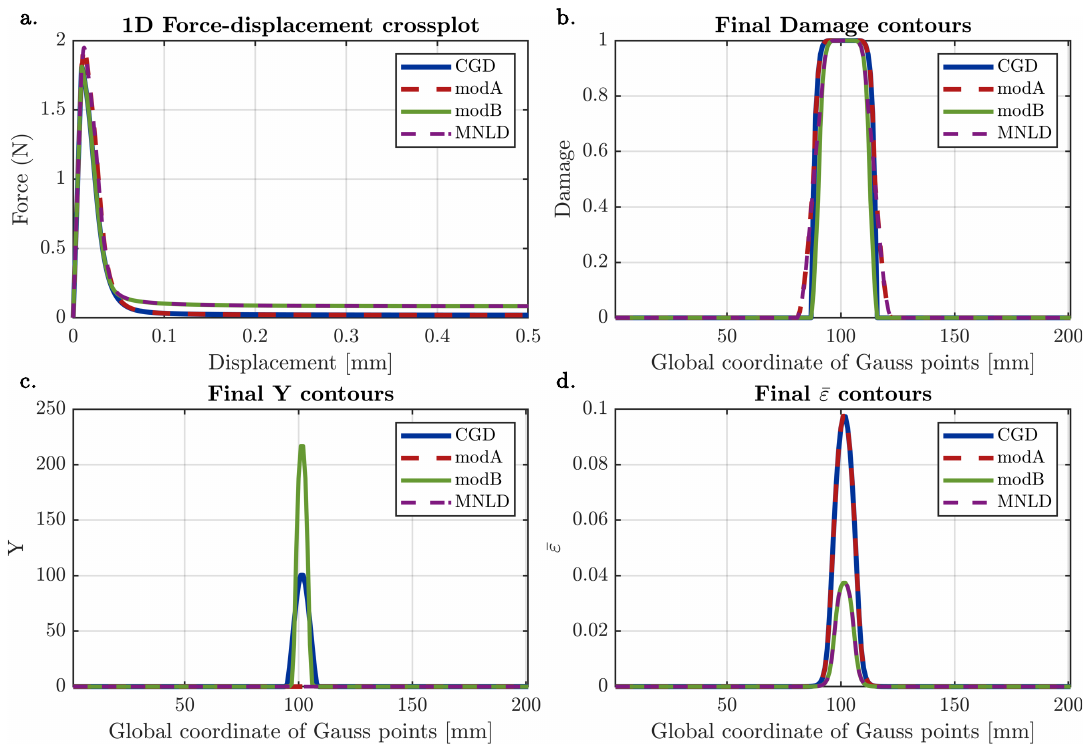}
    \end{minipage}
    \caption{Comparison of 1D benchmark results for CGD, modA, modB, and MNLD models. a) force–displacement curves, b) final damage contours, c) final $Y$ contours, and d) final non-local strain contours.}
    \label{fig:1D_contours}
\end{figure}

Fig. \ref{fig:1D_contours} displays the force–displacement response and the final increment contours of $d$, $Y$, and $\bar \varepsilon$  for the CGD, modA, modB, and MNLD models, respectively. The following remarks can be drawn based on these results: 

\begin{itemize}

    \item MNLD is the only model where both $Y$ becomes zero \emph{and} $\bar\varepsilon$ progressively decays in the region of the domain where damage reaches its maximum value. This provides the first strong evidence that the results are consistent with expectations mandated from the mathematical formulation.
    
    \item We observe differences both in the force-displacement and in the final damage contours across the four models. Here we need to emphasize that 1D damage problems are very unstable - once $d$ becomes 1, damage grows exponentially across the domain. The differences in these profiles will become more apparent and therefore will be more clearly appreciated when analyzing the response of 2D problems, as done in the following section.
      
\end{itemize}




\subsection{2D problems}
\label{NumEx: 2D problems}

In this subsection, three two-dimensional (2D) examples are presented to illustrate the results the MNLD model, and to examine the factors which influence its performance. For each case, a comparison with CGD is provided. The Single Notch Shear (SNS) problem serves as the primary example for evaluating MNLD, where the influence of the modified damage law parameters $s_1$ and $s_2$ is explored, and comparisons with the intermediate formulations modA and modB are presented. In the L-shaped panel example, we assess the sensitivity of MNLD to the maximum load step size ($\Delta \lambda_{max}$). Finally, the three-point bending example is analyzed, illustrating the robustness of the MNLD model through a widely adopted benchmark problem under flexural loading. In all problems, plane strain conditions are considered, and the decay function exponent is set as $n = 10^{2}$.

\subsubsection{Single Notch Shear (SNS) problem}
\label{NumEx:SNS}

\begin{figure}[ht]
    \centering
    \begin{subfigure}{8.2cm}
    \centering     
    \includegraphics[width=1\textwidth,trim = 11cm 4cm 11cm 4cm, clip]{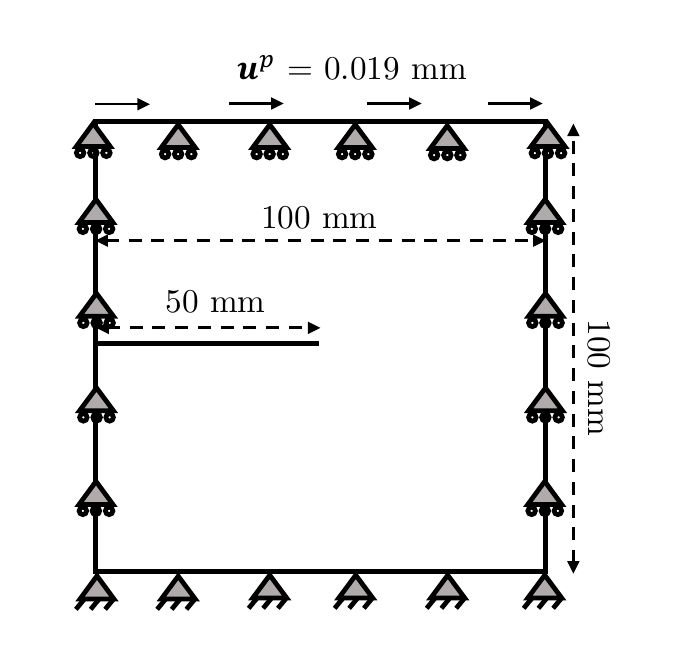}
    \caption{SNS schematic}
    \end{subfigure}
    \hfill
    \centering
    \begin{subfigure}{8.2cm}
    \centering     
    \includegraphics[width=1\textwidth,trim = 11.5cm 5cm 11.5cm 5cm, clip]{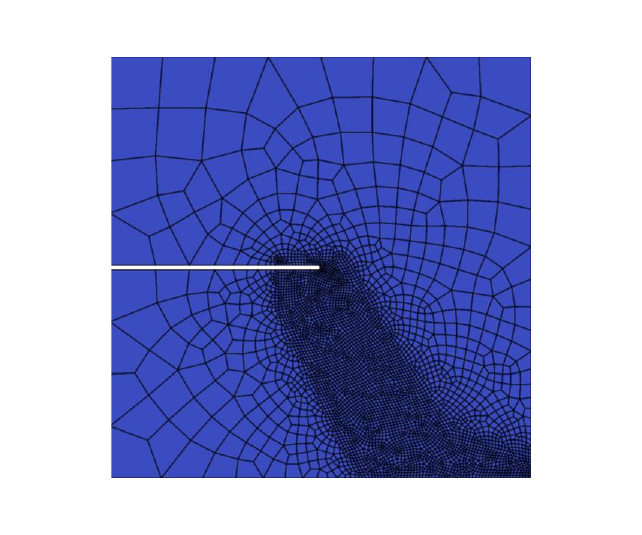}
    \caption{SNS Mesh}
    \end{subfigure}
    \caption{A schematic illustration of the (a) geometry and (b) mesh of the SNS problem. The finite element mesh is unstructured and comprises 4129 quadrilateral elements.}
    \label{fig:SNS Schematic & mesh}
\end{figure}

The first 2D example is the Single Notch Shear (SNS) problem \cite{pantidis2024fenn,treifi2009computations,zhou2018phase}. The specimen measures 100 mm × 100 mm with a horizontal notch. It is discretized using an unstructured mesh with 4129 elements, as shown in Fig \ref{fig:SNS Schematic & mesh}, with the element size in the refined zone being 0.75 mm × 0.75 mm, and the characteristic length $l_c$ set to 2 $mm$. A displacement-controlled shear load of $u^{p}$ = 0.019 mm is applied to the top surface of the domain, while the bottom surface is fully constrained, and roller boundary conditions are imposed on the remaining three surfaces. The shear modulus is $G = $ 125 MPa and Poisson's ratio is $\nu=$0.2. The modified Von Mises equivalent strain from Eqn. \eqref{deVree_strain_definition} and the modified Mazar's damage model from Eqn. \eqref{modified_Mazars_damage} are applied, with $\varepsilon_D = 10^{-4}$, $\alpha = 0.7$, $\beta = 10^{4}$, $s_1 = 1.5$ and $s_2 = 5.0$. The initial and maximum load-step sizes are $\Delta\lambda_0 = 10^{-4}$ and $\Delta\lambda_{max} = 5 \times 10^{-3}$, respectively.  

\subsubsection{Comparison with CGD}

\begin{figure}[H]
\centering
\includegraphics[width=1\textwidth,trim = 2.25cm 3cm 2.25cm 3cm, clip]{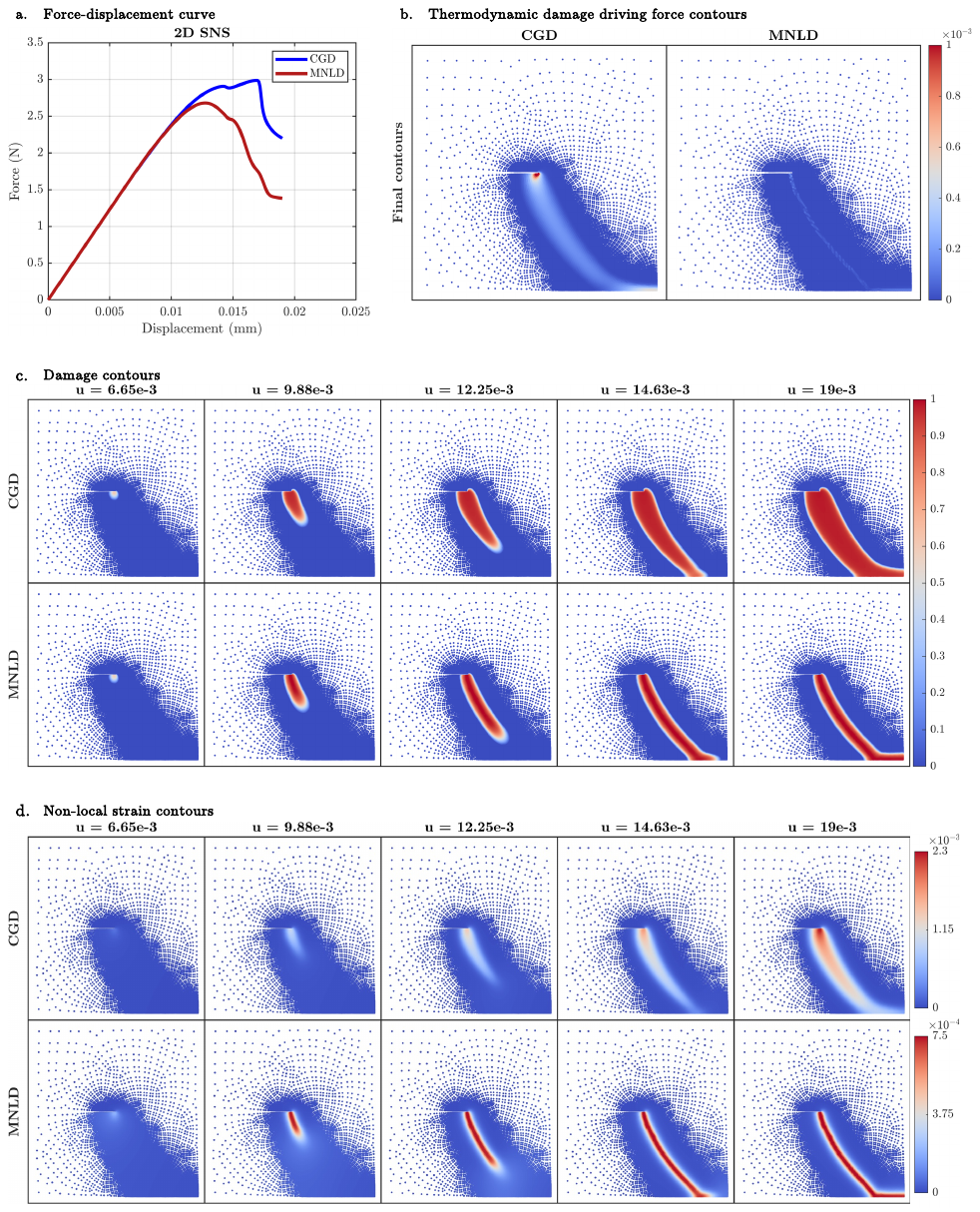}
\caption{Numerical results for the SNS test: comparison between the conventional gradient damage (CGD) and the modified non-local damage (MNLD) models. a) Force–displacement curves. b) Final thermodynamic damage driving force contours, showing a persistent, diffused non-zero field in CGD versus a near-zero field in MNLD. c) Damage distributions depicting a progressively widening damage band in CGD, vs fixed-width localization in MNLD. d) Non-local strain contours showing a broad, diffused strain evolution in CGD vs sharply localized non-local strain bands in MNLD. We emphasize the different limits in the colorbars of this subplot. All displacements are in mm.}
\label{SNS_contours}      
\end{figure}

The comparison presented in Fig. \ref{SNS_contours} highlights the markedly improved performance of MNLD over CGD. Fig. \ref{SNS_contours}a, demonstrates the force-displacement response of the two models. The $Y$ contours at the final increment, shown in Fig. \ref{SNS_contours}b, highlight the fundamental differences between the two models. The CGD model exhibits a persistent, diffused, non-zero $Y$ field even in fully damaged regions. In contrast, the MNLD model shows a narrow band of near-zero $Y$ along the crack path, indicating that the growth of $Y$ is effectively arrested in the fully damaged zone. A sequence of damage and $\bar\varepsilon$ contours at five displacement increments is shown in Fig. \ref{SNS_contours}c and Fig. \ref{SNS_contours}d, respectively. The following observations can be made from these figures:

\begin{itemize}
\item CGD: As the displacement increases, the damage and $\bar\varepsilon$ contours progressively widen, resulting in diffused and unrealistic profiles, which is consistent with well-documented results of the CGD model.

\item MNLD: Damage ($d$) and non-local equivalent strain ($\bar\varepsilon$) remain sharply localized with a fixed width as they propagate, even at higher displacement levels. This demonstrates MNLD’s effectiveness in maintaining a physically meaningful damage band and preventing spurious damage widening.
\end{itemize}

\subsubsection{Comparison against modA and modB}
\label{SNS_compare_modA_modB}

\begin{figure}[b!]
\centering
\includegraphics[width=1\textwidth,trim = 5.25cm 9cm 5.25cm 9cm, clip]{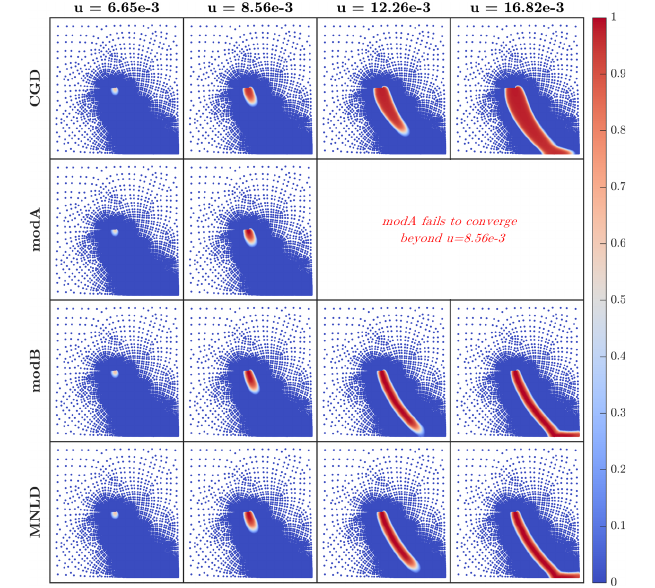}
\caption{Comparison of damage contours at four load increments for the CGD, modA, modB, and MNLD models. The CGD model exhibits progressively widening damage bands. The modA approach suffers from severe convergence issues even at low load levels. The modB model produces sharp, realistic contours but lacks thermodynamic consistency, compromising its reliability. The MNLD model combines the sharp localization achieved by modB with the thermodynamic consistency of modA, resulting in fixed-width damage evolution. All displacements are in mm.} 
\label{SNS_comparison_all_models}      
\end{figure}

\begin{figure}[t!]
\centering
\includegraphics[width=1\textwidth,trim = 2cm 6cm 0cm 0cm, clip]{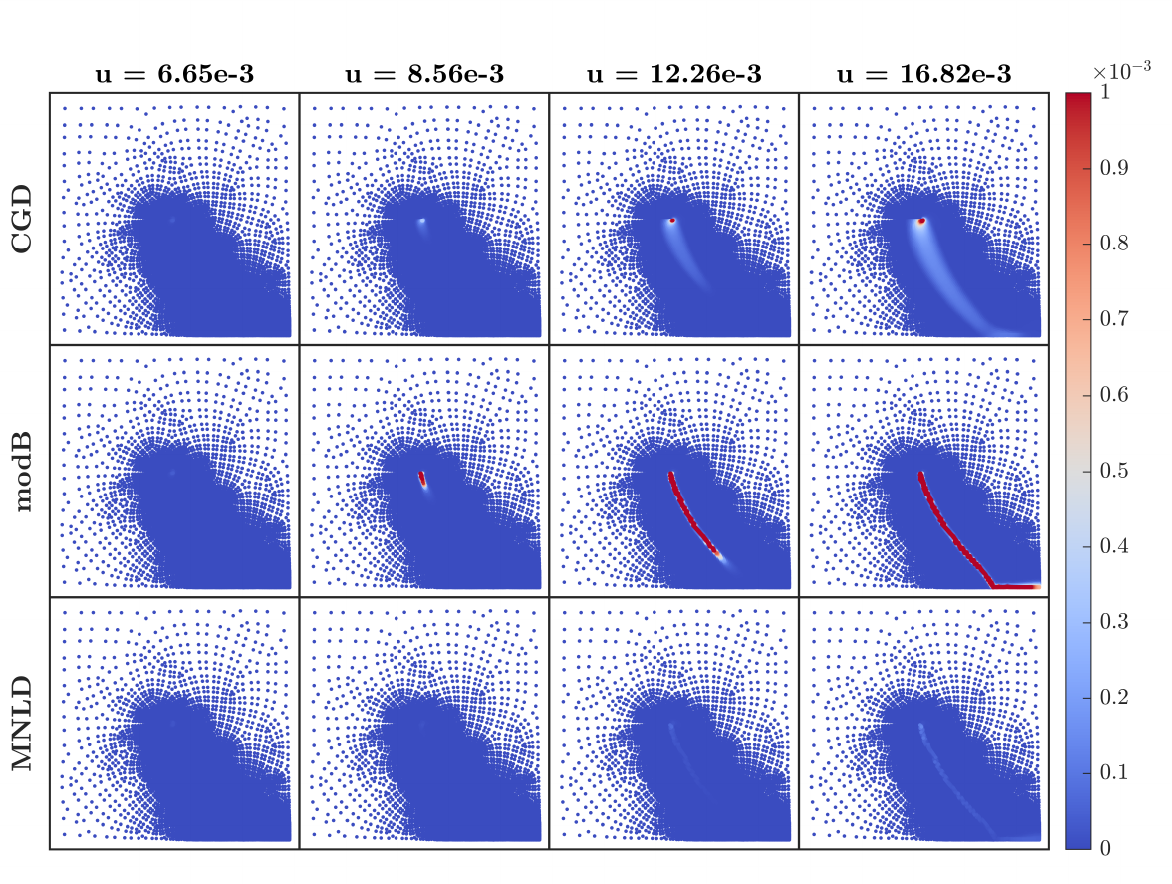}
\caption{Comparison of $Y$ contours at four load increments for the CGD, modB, and MNLD models. The $Y$ profile from CGD experiences the expected diffusion, while the modB analysis results in a localized profile with extremely high values ($\approx$ 2-3 orders of magnitude larger than the other models). MNLD shows the expected desired response, where $Y$ is narrow and near-zero across the full damage path. All displacements are in mm.} 
\label{SNS_Y_Contours}      
\end{figure}

In this subsection, we compare the damage contours at four displacement levels for modA and modB, with the corresponding results for CGD and MNLD included for ease of direct comparison. The results are demonstrated in Fig. \ref{SNS_comparison_all_models}. The modA model initially localizes damage, then suffers from severe convergence issues, failing to proceed beyond $u=8.56 \times 10^{-3}$ mm. The modB model exhibits sharp damage localization consistent with the desired fixed-width behavior; however, it has a non-zero $Y$ profile at maximum damage, as demonstrated in the 1D example and in Fig. \ref{SNS_Y_Contours}, demonstrating its thermodynamic inconsistency. This inconsistency limits its generalizability, compromising reliability and predictive accuracy despite visually favorable results in this particular example.

\subsubsection{Influence of modified damage model parameters $s_1$ and $s_2$}
\label{subsec:SNS_s1_s2_study}

\begin{figure}[H]
\centering
\includegraphics[width=1\textwidth,trim = 5.25cm 9cm 5.25cm 9cm, clip]{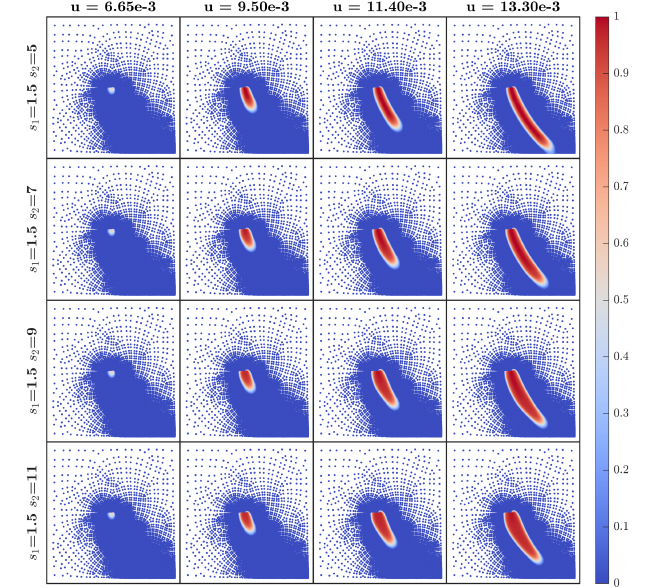}
\caption{Evolution of damage contours in the SNS problem at four load increments for the MNLD model with $s_1=1.5$ and $s2=\{5,7,9,11\}$ respectively. As $s_2$ increases, the enforcement of $d=1$ is activated later in the damage evolution regime, causing the damage contours to progressively resemble those of the conventional gradient damage (CGD) model by exhibiting wider damage bands. All displacements are in mm.} 
\label{SNS_s1s2_study_damage}      
\end{figure}

\begin{figure}[H]
\centering
\includegraphics[width=1\textwidth,trim = 5.25cm 9cm 5.25cm 9cm, clip]{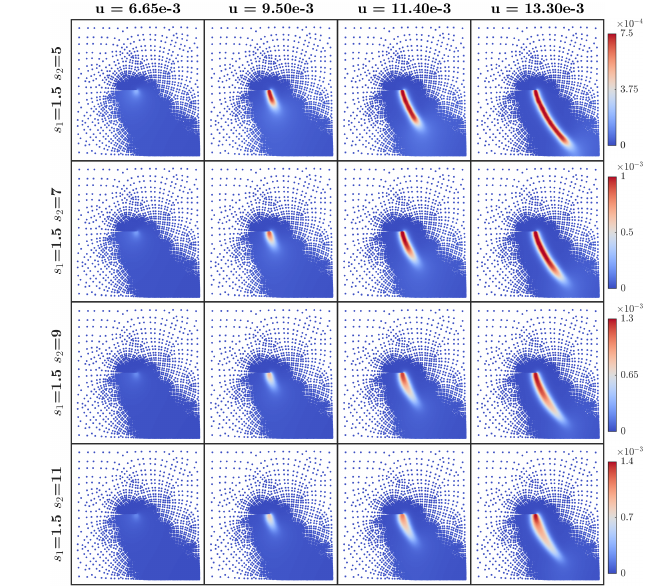}
\caption{Evolution of non-local strain contours in the SNS problem at four load increments for the MNLD model with $s_1 = 1.5$ and $s_2 = \{5,7,9,11\}$, respectively. As $s_2$ increases, the non-local strain field becomes increasingly diffused, with localization behavior gradually approaching that of the conventional gradient damage (CGD) model. All displacements are in mm.}

\label{SNS_s1s2_study_nonlocal_strain}      
\end{figure}

Next, we examine the role of the modified damage model parameters $s_1$ and $s_2$ on the behavior of the MNLD model. The evolution of damage and non-local strain contours for four cases with $s_1=1.5$ and $s_2=\{5,7,9,11\}$ are presented in Fig. \ref{SNS_s1s2_study_damage} and Fig. \ref{SNS_s1s2_study_nonlocal_strain}. At lower values of $s_2=\{5,7\}$, damage contours remain sharply localized with narrow, fixed-width bands propagating with increasing displacement load. Correspondingly, the $\bar\varepsilon$ fields exhibit confined bands that closely track the localized damage evolution. However, as $s_2$ increases, both damage and $\bar\varepsilon$ exhibit a progressively diffused behavior. This is because as $s_2$ increases, the transition to maximum damage is deferred further into the damage evolution regime (see Fig. \ref{fig:MazarsvsModifiedmazars}). Thus, the vanishing of $Y$ is delayed and the damage and $\bar\varepsilon$ behavior resembles that of the CGD model. 

\subsubsection{L-shaped panel}

\begin{figure}[H]
    \centering
    \begin{subfigure}{8.2cm}
    \centering     
    \includegraphics[width=1.6\textwidth,trim = 8cm 2cm 0cm 2cm, clip]{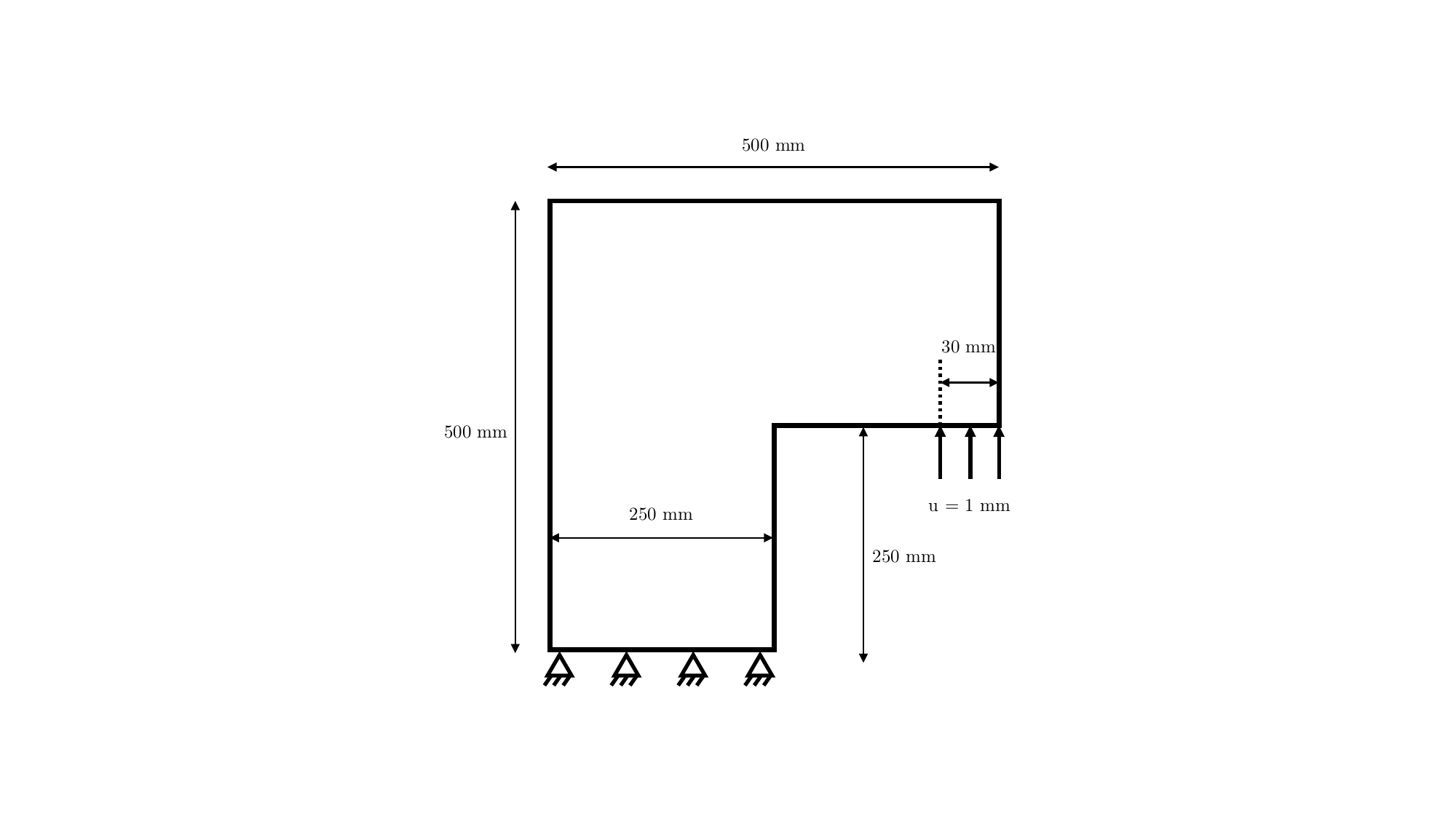}
    \caption{\footnotesize L-shaped panel schematic}
    \label{L_shaped_panel_schematic}
    \end{subfigure}
    \hfill
    \centering
    \begin{subfigure}{8.2cm}
    \centering     
    \includegraphics[width=1.8\textwidth,trim = 10cm 5cm 5cm 5cm, clip]{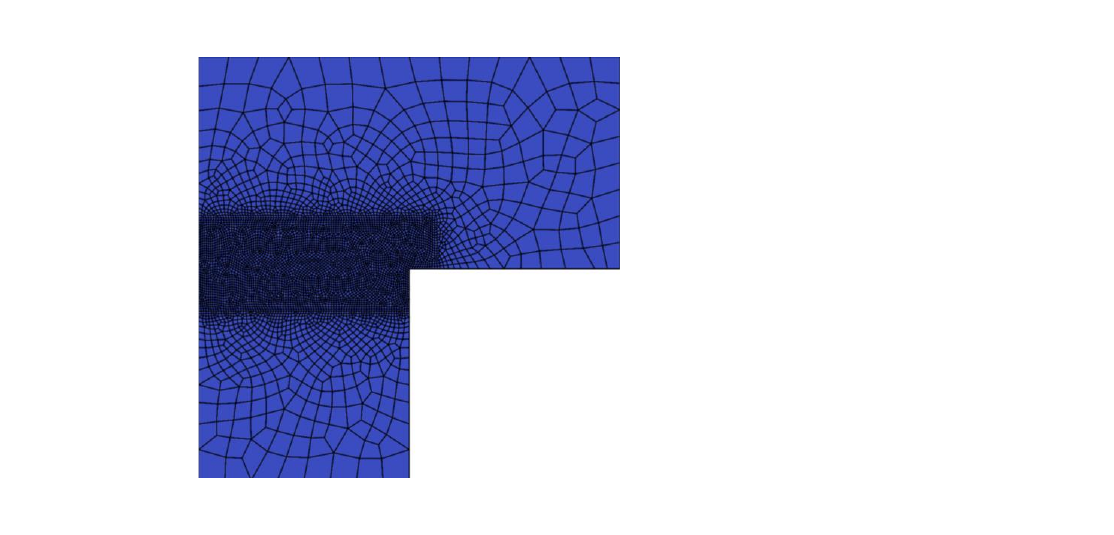}
    \caption{\footnotesize L-shaped pane mesh}
    \label{L_shaped_pane_mesh}
    \end{subfigure}
    \caption{A schematic illustration of the (a) geometry and (b) mesh of the L-shaped panel. The unstructured coarse mesh displayed here has 5398 quadrilateral elements and is used to model a problem with an $l_c$ = 6 mm.}
    \label{fig:L_shaped_schematic&mesh}
\end{figure}

The second 2D example is the standard L-shaped panel \cite{radulovic2011effective,huang2016efficient,winkler2004application}, aimed to evaluate the performance of MNLD at different $\Delta \lambda_{max}$ values. The dimensions of the domain are shown in Fig. \ref{L_shaped_panel_schematic}; a vertical displacement of $1$ mm is applied to a $30$ mm portion of the right end of the domain and the bottom surface is fully constrained. The unstructured mesh with 5398 quadrilateral elements, as depicted in Fig. \ref{L_shaped_pane_mesh}, has a refined region with element size $3$ mm $\times$ $3$ mm. The domain has $l_c=6$ mm, $G=$ 8000 MPa and $\nu=0.18$. The modified Von Mises equivalent strain definition (Eqn. \eqref{deVree_strain_definition}) and modified Geers damage model (Eqn. \eqref{modified_Geers_damage}) are used with $\varepsilon_D=2.5 \times 10^{-4}$, $\alpha=0.96$, $\beta=600$, $s_1=1.5$, and $s_2=9$. The initial and maximum load-step sizes are $\Delta\lambda_0=10^{-4}$ and $\Delta\lambda_{max}=4 \times 10^{-3}$, respectively. 

\begin{figure}[H]
\centering
\includegraphics[width=1\textwidth,trim = 2cm 3cm 2cm 3cm, clip]{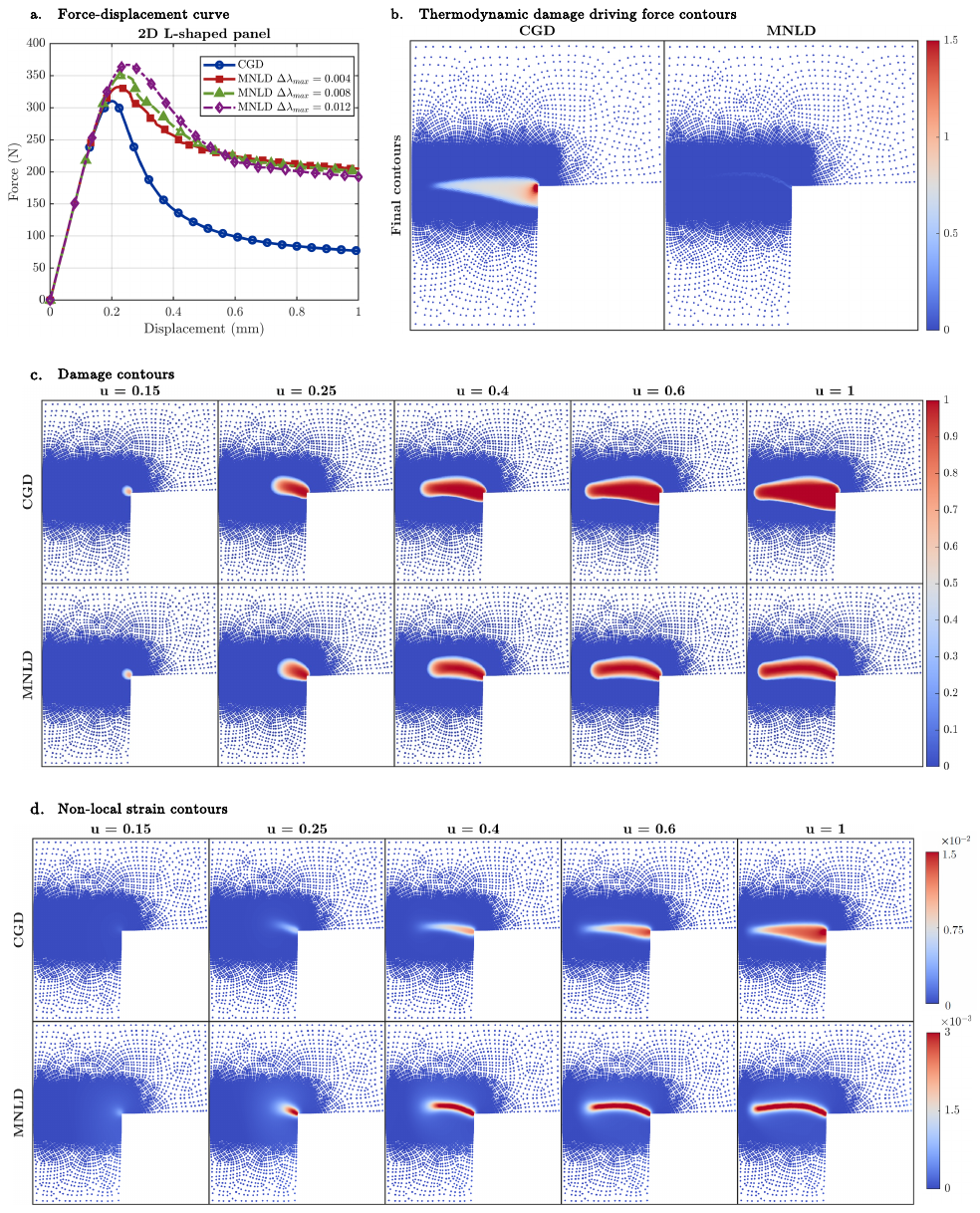}
\caption{Numerical results for the L-shaped panel test comparing the conventional gradient damage (CGD) and the modified non-local damage (MNLD) models. a) Force–displacement response of the CGD model compared against the MNLD model. b) Final $Y$ contours showing a non-zero diffused field in CGD and a vanishing field in MNLD. c) Damage contours illustrating a widening damage band in CGD and fixed-width localization in MNLD. d) Non-local strain contours revealing a highly diffused evolution in CGD and sharp localization in MNLD. All displacements are in mm.}
\label{L_shaped_contours}      
\end{figure}

Fig. \ref{L_shaped_contours} compares the performance of the MNLD and the CGD model. The force–displacement curves in Fig. \ref{L_shaped_contours}a, corresponding to different values of $\Delta \lambda_{\text{max}}$, exhibit differences in the softening and post-peak regions, with smaller $\Delta \lambda_{\text{max}}$ values having lower peak loads. The final $Y$ contours in Fig. \ref{L_shaped_contours}b show behavior similar to the SNS example: the CGD model exhibits a non-zero, spatially diffused driving force field, whereas the MNLD model demonstrates clear vanishing of the damage driving force in fully damaged regions. In Figs. \ref{L_shaped_contours}c and \ref{L_shaped_contours}d, the CGD model exhibits progressively widening damage and increasingly diffused $\bar\varepsilon$ fields with increasing displacement, whereas the MNLD model consistently maintains a narrow, fixed-width band in both damage and $\bar\varepsilon$. 

\begin{figure}[H]
\centering
\includegraphics[width=1\textwidth,trim = 3.5cm 9.5cm 3.5cm 9.5cm, clip]{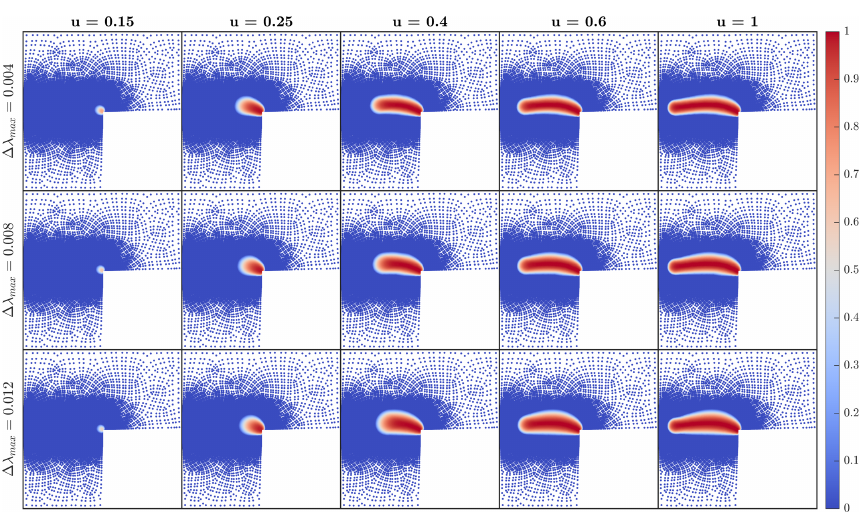}
\caption{Step size sensitivity study for the L-shaped panel problem using the MNLD model with step sizes of 4$\times10^{-3}$, 8$\times10^{-3}$, and 12$\times10^{-3}$. As the step size increases, the final damage band becomes progressively wider. This can be attributed to the growing influence of the auxiliary function $f(d,\prescript{k-1}{}{d})$ on stress, which consequently affects the damage evolution and strain localization behavior. All displacements are in mm.}
\label{L_shaped_dl_study}      
\end{figure}

To investigate the influence of $\Delta\lambda_{max}$, a study was conducted with $\Delta\lambda_{max} = 0.004$, 0.008, and 0.012, and the corresponding results are presented in Fig. \ref{L_shaped_dl_study}. The step-size sensitivity study reveals that as $\Delta\lambda_{max}$ increases in the MNLD model, the final damage bands become progressively wider. This widening indicates that larger step sizes amplify the effect of the auxiliary function $f_r$ on stress. Consequently, strain localization weakens, potentially compromising the accuracy of crack-path predictions and highlighting the importance of careful step-size selection for reliable simulations. This investigation supports our hypothesis presented in Section \ref{Sec:Methodology}, which states that the thermodynamic driving force $Y$ decays in the MNLD model as the step size becomes smaller, and consequently leads to narrower damage bands.

\subsubsection{Three-point bending (TPB)}
\begin{figure}[H]
    \centering
    \begin{subfigure}[b]{\textwidth}
        \centering
        \includegraphics[width=0.8\textwidth,trim = 5cm 7cm 5cm 7cm, clip]{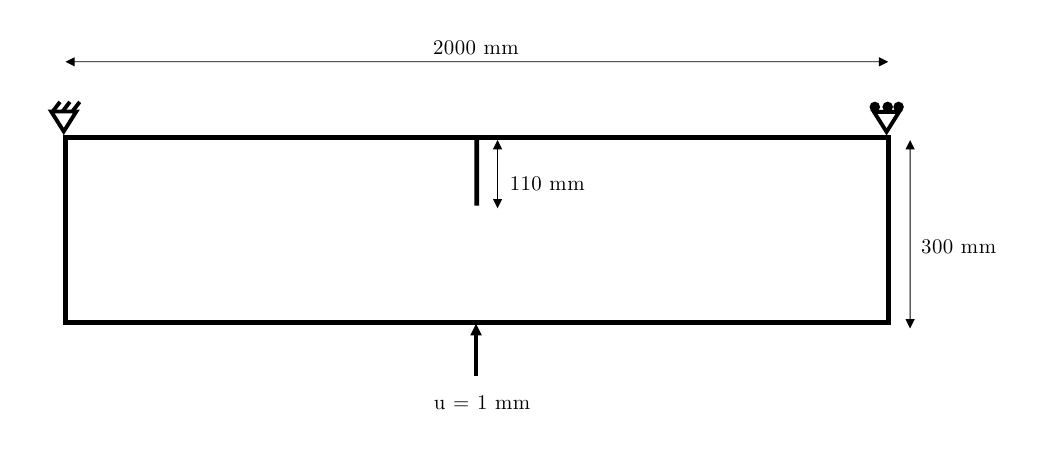} 
        \caption{\footnotesize Three point bending problem schematic}
        \label{fig:TPB_Schematic}
    \end{subfigure}
    
    \vspace{1cm} 
    
    \begin{subfigure}[b]{\textwidth}
        \centering
        \includegraphics[width=0.8\textwidth,trim = 9cm 10cm 8cm 10cm, clip]{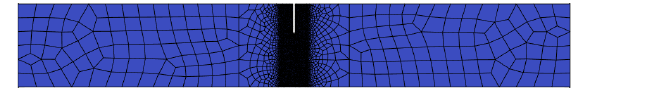} 
        \caption{\footnotesize Three point bending problem mesh}
        \label{fig:TPB_Mesh}
    \end{subfigure}
\caption{A schematic illustration of the a) geometry and b) mesh of the three point bending problem. The unstructured mesh displayed here has 5029 elements and a $l_c = 6 mm$ is used.}
\label{TPB_Schematic&Mesh}
\end{figure}

The final 2D benchmark problem is the Three-Point Bending (TPB) test \cite{de2016gradient,askes2000dispersion}, aimed at assessing the robustness of MNLD under flexural loading. The TPB specimen is modeled as a 2000 mm × 300 mm domain with a centrally located notch of length $110$ mm, discretized using an unstructured mesh containing 5029 elements, as depicted in Fig. \ref{TPB_Schematic&Mesh}. Elements in the refined region have dimensions of $3$ mm × $3$ mm, and $l_c=6$ mm. A displacement-controlled load of $1$ mm is applied at the midpoint of the bottom edge, with the supports fully constraining the top left corner and rollers applied on the top right corner to ensure stability. The material properties are $G=8333.4$ MPa and $\nu=0.2$. The modified Mazars equivalent strain (Eqn. \eqref{Mazars_Equivalent_strain_tension}) and modified Geers damage law (Eqn. \eqref{modified_Geers_damage}) are applied with $\varepsilon_D=10^{-4}$, $\alpha=0.99$, $\beta=1500$, $s_1=1.5$, and $s_2=8$. Load-step sizes are initialized with $\Delta\lambda_0=10^{-4}$ and capped at $\Delta\lambda_{max}=10^{-3}$.

\begin{figure}[H]
\centering
\includegraphics[width=1\textwidth,trim = 2cm 3cm 2cm 3cm, clip]{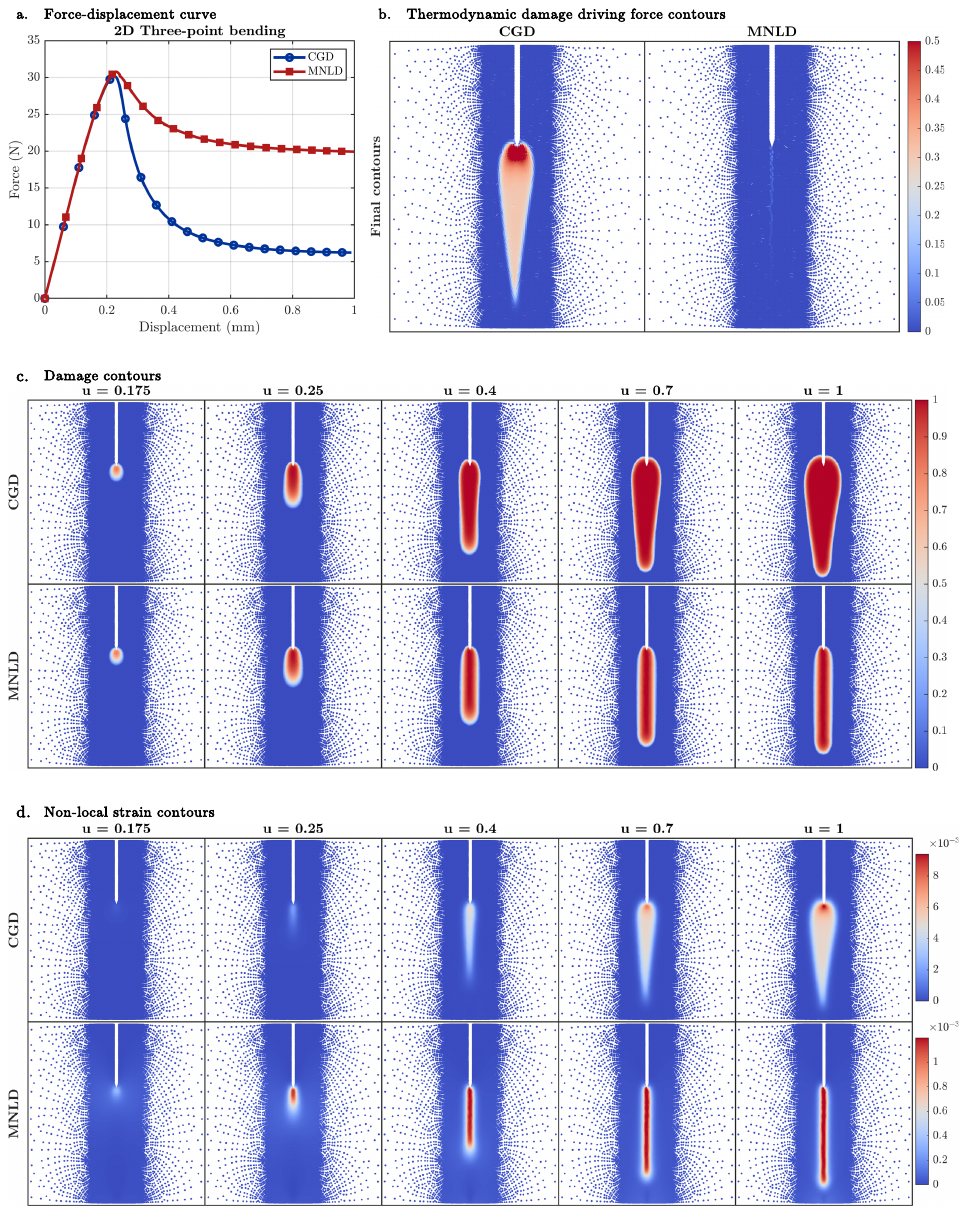}
\caption{ Comparison of CGD and MNLD models in a 2D three-point bending problem. a) Force–displacement curves show a softer response with an earlier force drop in the CGD model, while the MNLD model sustains a longer post-peak response b)  Final $Y$ contours highlight a diffused non-zero field in CGD and a nearly invisible field in MNLD c) Damage evolution contours at five displacement levels {u = 0.175, 0.25, 0.4, 0.7, 1} illustrate progressive widening of the damage band in CGD, whereas MNLD maintains a consistent narrow damage band. d) Non-local strain contours at the same displacement levels reveal highly diffused fields in CGD compared to sharply localized ones in MNLD. All displacements are in mm}
\label{TPB_contours}      
\end{figure}

In Fig. \ref{TPB_contours}, we compare the performance of CGD with MNLD in the 2D three point bending problem. The force-displacement curves in Fig. \ref{TPB_contours}a show the MNLD model having a gradual post-peak decline unlike CGD that drops sharply. The $Y$ contours of the MNLD model in Fig. \ref{TPB_contours}b are scarcely discernible, in stark contrast to the CGD model, which exhibits highly diffused contours indicative of persistent spurious energy release even after complete damage. 
The damage ($d$) and non-local equivalent strain ($\bar\varepsilon$) contours in \ref{TPB_contours}c and \ref{TPB_contours}d, respectively, highlight again the stark differences between the two models as the displacement load increases. Similar to the L-shaped panel and SNS problems, the MNLD model in this case also demonstrates excellent field localization, consistently maintaining a fixed-width damage band across all load levels. This behavior validates its capability to confine damage to a physically meaningful scale.

Overall, the numerical results presented in this section demonstrate the different aspects of MNLD and its performance in comparison to CGD, modA and modB. The following are some key highlights:

\begin{itemize}
    \item The MNLD model successfully mitigates the spurious damage widening observed in CGD at the wake of the damage band
    \item In the MNLD formulation, $Y$ vanishes as damage reaches 1
    \item The non-local strain ($\bar\varepsilon$), which is implicitly related to the forcing term $f_r\varepsilon$, progressively decays as damage nears its maximum value.
\end{itemize}

\section{Summary and conclusions}
\label{Sec:Conclusion}

In this work we introduce a modified non-local damage (MNLD) model designed to overcome the persistent issue of widening damage bands that is observed in conventional gradient damage models. The proposed MNLD formulation enforces two critical conditions: a) the thermodynamic damage driving force $Y$ vanishes as damage reaches its maximum, and b) the forcing term $f_r\varepsilon$ decays correspondingly. This setup ensures that the width of the damage band remains fixed and physically meaningful throughout the evolution of failure. 

A comprehensive suite of benchmark problems, including the 1D bar under tension, Single Notch Shear (SNS) specimen, the L-shaped panel test, and a 2D three-point bending problem, were examined to evaluate the model’s performance and test its robustness under different loading scenarios. In the 1D example, the MNLD model resulted in a complete absence of $Y$ in regions where damage evolution had concluded, while the evolution of $\bar\varepsilon$ was significantly suppressed compared to the CGD model—providing early verification of the model’s intended behavior. 
Across all 2D examples, the MNLD model consistently maintained a narrow fixed-width damage band, in stark contrast to the CGD model, which exhibited widening damage zones with increasing loads. The influence of the modified damage parameters ($s_1$, $s_2$) and the maximum load step size ($\Delta \lambda_{\mathrm{max}}$) was also investigated to better elucidate the conditions required for the optimal performance of the MNLD model.

These results clearly indicate that the MNLD approach effectively resolves the persistent challenge of spurious damage widening in gradient-enhanced damage models, while avoiding the complications of the localizing gradient models. The straightforward implementation into existing gradient damage frameworks, combined with its thermodynamic consistency and fixed-width damage behavior, makes the MNLD model a promising candidate for reliable failure prediction in brittle and quasi-brittle materials.

\section*{Data availability}
\label{Sec:Data availability}
The code and datasets used in this work will be made publicly available upon publication of the article.  

\section*{Acknowledgments}
\label{Sec:Acknowledgements}
This work was partially supported by the Sand Hazards and Opportunities for Resilience, Energy, and Sustainability (SHORES) Center, funded by Tamkeen under the NYUAD Research Institute Award CG013. The authors also acknowledge the support of the NYUAD Center for Research Computing, which provided resources, services, and staff expertise that contributed to this work.

\bibliography{bibliography}
\newpage
\appendix




\section{Derivation of the governing equations for the MNLD model}
\label{Apx:psi_defination}

Here we present the derivation of the governing equations for the Modified non-local Damage (MNLD) model. The Helmholtz free energy density is defined as:
\begin{equation}
\psi (\varepsilon_{ij},\bar\varepsilon,d) = \dfrac{1}{2} [(1-d) + f_a] \varepsilon_{ij} C_{ijkl} \varepsilon_{kl} + \dfrac{1}{2}h \left[({f_r \varepsilon} - \bar\varepsilon)^2 + c \bar\varepsilon_{,i}^2 \right]
\end{equation}

The derivatives of the free energy with respect to the state variables $d$,$\varepsilon_{ij}$ and $\bar\varepsilon$ are:
\begin{equation}
\dfrac{\partial \psi}{\partial \varepsilon_{ij}} =  [(1-d) + f_a] C_{ijkl} \varepsilon_{kl} + h(f_r \varepsilon - \bar\varepsilon) \dfrac{\partial \varepsilon}{\partial \varepsilon_{ij}}
\end{equation}

\begin{equation}
\dfrac{\partial \psi}{\partial \bar\varepsilon} =
 - h ((f_r\varepsilon - \bar\epsilon)
+
h c \bar\varepsilon_{,i} 
\end{equation}

\begin{equation}
\frac{\partial \psi}{\partial d} = \frac{1}{2} \left[-1 + \dfrac{\partial f_a}{\partial d} \right] \varepsilon_{ij} C_{ijkl} \varepsilon_{kl} - h \left( f_r\varepsilon - \bar{\varepsilon} \right) \varepsilon
\end{equation}

\begin{equation}
Y = -\frac{\partial \psi}{\partial d} = \frac{1}{2} \left[1 - \dfrac{\partial f_a}{\partial d} \right] \varepsilon_{ij} C_{ijkl} \varepsilon_{kl} + h \left( f_r\varepsilon - \bar{\varepsilon} \right) \varepsilon \dfrac{\partial f_r}{\partial d}
\end{equation}

The term $\dfrac{\partial f_r}{\partial d}$ becomes zero for large values of $n$ and $d<1$. For thermodynamic consistency, the following global dissipation inequality must be satisfied:

\begin{equation}
\dot{\mathscr{D}} = \int_V (\sigma_{ij} \dot{\varepsilon_{ij}} - \dot{\psi}) dV \geq 0
\end{equation}

\begin{align}
\dot{\mathscr{D}} & = \int_V \left(\sigma_{ij} \dot{\varepsilon_{ij}} 
-
\left[
\dfrac{\partial \psi}{\partial \varepsilon_{ij}} \dfrac{\partial \varepsilon_{ij}}{\partial t} 
+ 
\dfrac{\partial \psi}{\partial d} \dfrac{\partial d}{\partial t} 
+
\dfrac{\partial \psi}{\partial \bar\epsilon} \dfrac{\partial \bar\epsilon}{\partial t} 
\right]
\right) dV  \\
& =
\int_V \underbrace{\sigma_{ij} \dot{\varepsilon_{ij}} 
- 
\dot{\varepsilon_{ij}} \left[[(1-d)+ f_a] C_{ijkl} \varepsilon_{kl} + h(f_r \varepsilon - \bar\varepsilon) \dfrac{\partial \varepsilon}{\partial \varepsilon_{ij}}\right]}_{\dot{\mathscr{D}_{\varepsilon_{ij}}}} \\
& \underbrace{+ \dot{d} \left[ \frac{1}{2} \left[-1 + \dfrac{\partial f_a}{\partial d} \right]\varepsilon_{ij} C_{ijkl} \varepsilon_{kl} - h \left( f_r\varepsilon - \bar{\varepsilon} \right) \varepsilon \right]}_{\dot{\mathscr{D}_{d}}}
\underbrace{+ \dot{\bar\varepsilon} \left[
 - h (f_r\varepsilon - \bar\epsilon)
+
h c \bar\varepsilon_{,i}   \right] }_{\dot{\mathscr{D}_{\bar\varepsilon}}} dV
\end{align}

\begin{equation}
\dot{\mathscr{D}} = \dot{\mathscr{D}_{\varepsilon_{ij}}} + \dot{\mathscr{D}_{d}} + \dot{\mathscr{D}_{\bar\varepsilon}} \geq 0
\end{equation}

The governing equations are obtained from the non-dissipative terms $\dot{\mathscr{D}_{\varepsilon_{ij}}} $ and $\dot{\mathscr{D}_{\bar\varepsilon}}$.

\subsection{$\dot{\mathscr{D}_{\varepsilon_{ij}}}$}
\begin{align}
\dot{\mathscr{D}_{\varepsilon_{ij}}} & = 0
\end{align}

\begin{align}
\int_V \dot{\varepsilon_{ij}} \left[\sigma_{ij}  
- 
 \left[[(1-d)+ f_a] C_{ijkl} \varepsilon_{kl} + h(f_r \varepsilon - \bar\varepsilon) \dfrac{\partial \varepsilon}{\partial \varepsilon_{ij}}\right] \right] \ dV = 0
\end{align}

Thus, 

\begin{equation}
\sigma_{ij}  
 = 
 \left[[(1-d)+ f_a] C_{ijkl} \varepsilon_{kl} + h(f_r \varepsilon - \bar\varepsilon) \dfrac{\partial \varepsilon}{\partial \varepsilon_{ij}}\right]
 \label{New_stress_definition}
\end{equation}

The expression in Eqn. \ref{New_stress_definition} represents the new definition of stress.

\subsection{$\dot{\mathscr{D}_{\bar\varepsilon}}$}

\begin{equation}
\dot{\mathscr{D}_{\bar\varepsilon}} = 0
\end{equation}

\begin{equation}
\int_V \dot{\bar\varepsilon} \left[
 - h (f_r\varepsilon - \bar\epsilon)
+
h c \bar\varepsilon_{,i}   \right] \ dV  = 0
\end{equation}

\begin{equation}
\int_V \dot{\bar\varepsilon} \left[
h (f_r\varepsilon - \bar\epsilon)
+
h c \bar\varepsilon_{,ii}  \right] \ dV  - \int_S - \dot{\bar\varepsilon} h c \bar\varepsilon_{,i} n_i\ dS= 0
\end{equation}

Following earlier derivations of non-local gradient relationships (e.g. \cite{peerlings2004thermodynamically, mobasher2018thermodynamic}), the localization of the above equation leads to:
\begin{equation}
\left[
 (f_r\varepsilon - \bar\epsilon)
+
 c \bar\varepsilon_{,ii}  \right] = 0~~  \text{in}~ V 
\end{equation}

\begin{equation}
\varepsilon_{,i} n_i = 0~~  \text{on}~ S
\end{equation}

\end{document}